\documentclass[manuscript]{aastex}
\usepackage{tikz}
\usepackage{amsmath}

\usepackage{graphicx}

\usepackage{pdflscape}

\slugcomment{\textbf{The Astronomical Journal, in press} }

\shorttitle{Fragments of 331P}
\shortauthors{Jewitt}

\begin{document}

\title{Fragmenting Active Asteroid 331P/Gibbs }


\author{David Jewitt$^{1,2}$, Jing Li$^1$ and Yoonyoung Kim$^3$\\
} 
\affil{$^1$Department of Earth, Planetary and Space Sciences,
UCLA, 
595 Charles Young Drive East, 
Los Angeles, CA 90095-1567\\
$^2$Department of Physics and Astronomy,
University of California at Los Angeles, \\
430 Portola Plaza, Box 951547,
Los Angeles, CA 90095-1547\\
$^3$Institut for Geophysik und Extraterrestrische Physik, Technische Universitat Braunschweig,
Mendelssohnstr. 3, 38106 Braunschweig, Germany
}

\email{jewitt@ucla.edu}

\begin{abstract}

We describe active asteroid 331P/Gibbs (2012 F5) using archival Hubble Space Telescope data taken between 2015 and 2018. 331P is an outer main-belt active asteroid with a long-lived debris trail that formed in 2011. Embedded in the debris trail we identify 19 fragments with radii between 0.04 and 0.11 km (albedo 0.05 assumed) containing about 1\% of the mass of the primary nucleus. The largest shows a photometric range ($\sim$1.5 magnitudes), a V-shaped minimum and a two-peaked lightcurve period near 9 hours, consistent with a symmetric contact binary (Drahus and Waniak 2016). Less convincing explanations are that 331P-A is a monolithic, elongated splinter or that its surface shows hemispheric 4:1 albedo variations. The debris trail is composed of centimeter sized and larger particles ejected with characteristic 10 cm s$^{-1}$ speeds following a size distribution with index $q$ = 3.7$\pm$0.1 to 4.1$\pm$0.2. The HST data show that earlier, ground-based measurements of the nucleus were contaminated by near-nucleus debris, which cleared by 2015. We find that the primary nucleus has effective radius 0.8$\pm$0.1 km and is in rapid rotation (3.26$\pm$0.01 hour, range 0.25 magnitudes, minimum density 1600 kg m$^{-3}$ if strengthless. The properties of 331P are consistent with a) formation about 1.5 Myr ago by impact shattering of a precursor body (Novakovic et al. 2014) b) spin-up by radiation torques to critical rotation c) ejection of about 1\% of the nucleus mass in mid-2011 by rotational instability and d) subsequent evolution of the fragments and dispersal of the debris by radiation pressure.

\end{abstract}


\section{INTRODUCTION}
\label{intro}
331P/Gibbs (2012 F5) (hereafter 331P) was discovered on UT 2012 March 22 (Gibbs et al.~2012).  The object showed both the physical attributes of a comet (in the form of a long, thin dust tail) and the orbital elements of an asteroid.  The orbital semimajor axis, eccentricity and inclination are 3.005 AU, 0.042 and 9.7\degr, respectively, corresponding to a Tisserand parameter with respect to Jupiter, $T_J$ = 3.228.   This is significantly above the nominal $T_J$ = 3 dividing line  separating asteroids from comets (Vaghi 1973), and also above the practical $T_J$ = 3.08 line  used to distinguish active asteroids from short-period comets (Jewitt 2012, c.f.~Hsieh and Haghighipour 2016).  Numerical simulations indicate that 331P is dynamically stable on 10$^9$ year timescales (Stevenson et al.~2012). It is simply a main-belt asteroid that ejects observable quantities of dust. 

As a new member of the exciting and still poorly understood active asteroids population, 331P attracted immediate observational attention.  Moreno et al.~(2012) and Stevenson et al.~(2012) used the position angle and morphology of the tail to infer an impulsive origin on UT 2011 July 1$\pm$11  and UT 2011 July 7$\pm$20, respectively.  As was the case with the impact-caused active asteroid P/2010 A2 (Jewitt et al.~2010), the $\sim$8 month delay between the release of the dust and the detection of the object is at least partly explained by the  small solar elongation (which reached a minimum $\sim$2\degr~in 2011 August and remained $<$90\degr~until 2011 December).    

A statistically significant mini-cluster of eight  asteroids dynamically associated with 331P was identified by Novakovi{\'c} et al.~(2014), suggesting formation by impact fragmentation of a parent body. From the tight convergence of the orbital elements within this cluster, these authors estimated a separation age of only 1.5$\pm$0.1 Myr.   331P is small compared to the largest  cluster member, 20674 (1999 VT1), which has diameter $\sim$18 km (assuming the same albedo), and the parent body may have been $\sim$24 km in diameter (Novakovi{\'c} et al.~2014).  Deep imaging observations taken in 2014  showed that 331P is rotating with a photometric range $\sim$0.2 magnitudes and a two-peaked period $P = 3.24\pm0.01$ hour, and revealed four fragments embedded in the dust tail (Drahus et al.~2015).  The short rotational period of 331P suggests a potential role for rotational breakup, instead of impact as the origin of the comet-like appearance.  

The available observations thus reveal 331P as an intriguing object.   The connection between its formation as an impact fragment $\sim$1.5 Myr ago and the modern episode of mass loss is unclear.  Did 331P experience another impact to eject material in 2011, or is rotational instability to be implicated? The outer-belt orbit of 331P is similar to the orbits of several active asteroids in which the mass loss is  likely due to sublimation, notably the prototype main-belt comet  133P/Elst-Pizarro (7968) (Hsieh and Jewitt 2006),  raising the possibility that sublimation might also play a role in 331P. The ambiguities suggested by these questions highlight the difficulties encountered in the interpretation of the active asteroids as a population.

In this paper we examine archival Hubble Space Telescope (HST) observations  taken under General Observer programs 14192, 14475, 14798 and 15360 (PI: M. Drahus) in order to characterize this intriguing object.  These data have not been published other than as a brief meeting abstract (Drahus and Waniak 2016).  We first  examine the physical properties of 331P and its ejecta as revealed at HST resolution.  In a separate work, we will analyse the dynamics of the fragments.

\section{OBSERVATIONS}

\subsection{ Photometry and General Appearance}
The WFC3 camera consists of two 2k$\times$4k charge coupled devices separated by a 1.2\arcsec~gap (Dressel 2012).  The image scale is 0.04\arcsec~pixel$^{-1}$, giving a Nyquist-sampled resolution near 0.08\arcsec, corresponding to $\sim$120 km at 2 AU geocentric distance.   The practical  limits to HST point-source photometry  are set by cosmic rays, which pepper the detector in great abundance.  Our philosophy is to avoid heroic efforts to recover images photometrically compromised by cosmic rays. We simply exclude severely compromised images from further consideration and attempt only light editing (by interpolation of surrounding pixels) of images showing minor cosmic ray and image defect interference.   A small minority of the images were so affected.

\begin{figure}
\epsscale{0.99}
%\plotone{SB.pdf}
%\plotone{images.pdf}
%\plotone{dec28.pdf}
\plotone{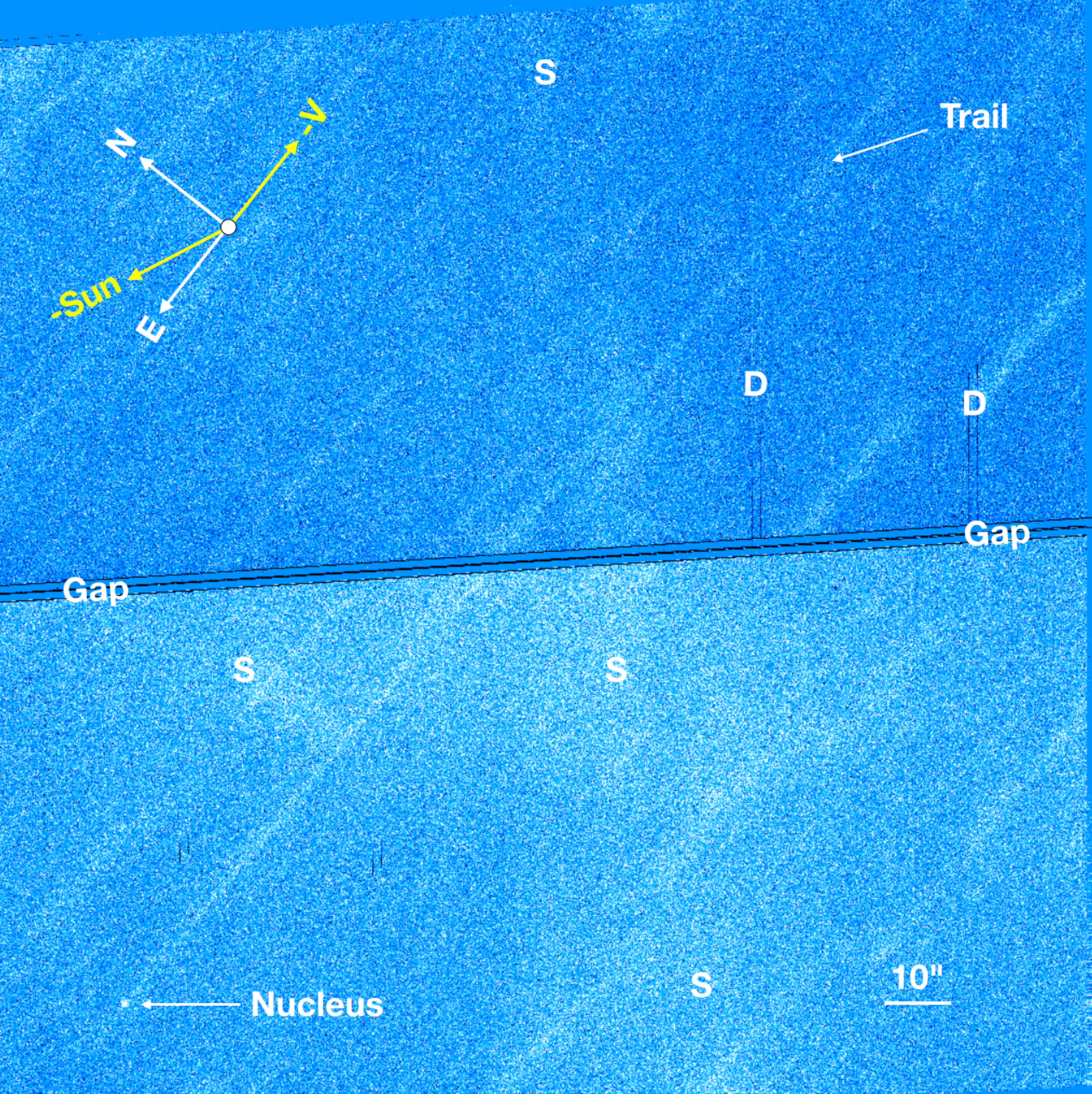}

\caption{Composite full-frame  (4k$\times$4k pixel) image from UT 2015 December 28 to show the impact of numerous imperfectly removed  field stars and galaxies, diffuse scattered light in the camera  (S),  various CCD defects (D) and the chip gap.  The cardinal directions are marked in the upper left, as are the projected anti-solar direction (-Sun) and the negative heliocentric velocity vector (-V).  A 10\arcsec~scale bar is shown.  \label{dec28}}
\end{figure}

Limits to the photometry of diffuse sources are set by internally scattered light from bright objects both inside and outside the field (c.f.~Figure \ref{dec28}), with particular impact on measurements of the diffuse trail.  In addition to scattered light, background stars and galaxies are smeared by strong parallactic motion on solar system sources.  When making composite images, we experimented with different rejection schemes to reduce these background variations.  Smeared galaxies are particularly difficult to suppress, because they are of large angular extent and frequently overlap in successive images.  Example residuals are evident in the diagonal streaks in Figure (\ref{dec28}).  In several of the data sets, the residuals caused by scattered light and trailed field objects are comparable to or brighter than the surface brightness of the 331P dust trail, making accurate measurement particularly difficult.    

We relied on the use of small photometry apertures to reduce the impact of both cosmic rays and background variations.  For the discrete fragments, most of which are  faint and highly susceptible to cosmic ray and other background uncertainties, we used a photometry aperture of 5 pixels (0.2\arcsec)~radius with sky determined from the median signal within a contiguous annulus extending to 4.2\arcsec.  The use of a relatively large sky annulus was found experimentally to give a more stable and accurate estimate of the local sky.  For 331P itself, we  experimented with nested apertures  5, 7 and 9 pixels (0.20\arcsec, 0.28\arcsec and 0.36\arcsec) in projected radius with sky subtraction from a contiguous annulus extending to 4.2\arcsec.  We find that the signal between 0.20\arcsec~and 0.36\arcsec~is about 8\% of the signal in the 0.2\arcsec~radius aperture, and that this signal fraction does not vary with variations in the central magnitude.  The sacrifice of a small fraction of the total signal from each object is warranted by the improved reduction of noise and rejection of interfering cosmic rays and image defects, and does not affect the relative photometry between fragments.  We have not made a correction for the 8\% reduction in photometrically estimated cross-sections (4\% in fragment radii); these corrections are, in any case, inconsequential compared to the much larger uncertainties introduced by the unknown albedos and phase functions of  the target bodies.  Our use of small apertures is only possible because of the remarkable stability of the  point-spread function of HST (Dressel 2012).

%\subsection{General Appearance}

We found useful observations of 331P taken in eight epochs spread over the period from late 2015 to mid-2018, using a total of 30 HST orbits.  
We did not find useful images from visits on UT 2018 May 4 and 5, and 2018 June 2, despite the fact that the ephemeris was no less accurate on these dates, and 331P could not have been faint enough to elude detection by HST.  Data from these dates are not further discussed here.  A journal of observations is given in Table (\ref{geometry}).

Figure (\ref{keycode}) shows  image composites from each epoch of observation, all rotated to bring the central axis of the trail to the horizontal.  The images have been scaled to a common geocentric distance, $\Delta$ = 2.011 AU (i.e.~the distance of the first observation on UT 2015 December 25), and co-aligned on the primary nucleus labeled 331P.    The panels are variously affected by imperfect removal of scattered light and trailed field objects, with the most glaring residuals apparent on the right hand (western) side of the panel for UT 2017 Mar 08.  Nevertheless, a consistent appearance is evident in the data from 2015 to 2018: 331P shows a bright primary nucleus, with a faint debris trail to the west in which are peppered point source fragments.  The images also show ultrafaint diffuse material to the east of the primary.  

Fragments 331P-A, -B and -C are prominent in each panel and are labeled in red in order to guide the eye.  Other fragments are identified only in the panel where they are first apparent and discussed in Section \ref{fragments}.  Note that the sky-plane positions of the fragments in Figure (\ref{keycode}) are affected by the changing phase angle of observation so that a steady expansion from the primary cannot be discerned.

\subsection{Primary Nucleus}
\label{primary}
We computed  the mean absolute magnitude of 331P   from the apparent V-band magnitude, $V$, using

\begin{equation}
H = V - 5\log_{10}(r_H \Delta) + 2.5\log_{10}(\Phi(\alpha))
\label{HV}
\end{equation}

\noindent where   $r_H$ and $\Delta$ are the heliocentric and geocentric distances in AU, respectively, and $\Phi(\alpha)$ is the phase function, equal to the ratio of the brightness measured at phase angle $\alpha$ to that at $\alpha$ = 0.  The phase function is unmeasured in 331P and is not necessarily the same for dust as for macroscopic objects.  Based on measurements of asteroids we adopt $2.5\log_{10}(\Phi(\alpha)) = -0.04\alpha$ (Martikainen et al.~2021) and note that a $\pm$50\% error in this value would affect the derived absolute magnitudes by up to $\pm$0.4 magnitudes.
.   

The absolute magnitudes are plotted as a function of the date of observation in Figure (\ref{absmags}), where we include both data from HST and from the literature.  Measurements taken in the R filter were converted to V using V-R = 0.4$\pm$0.1 (Novakovic et al.~2014). Measurements from the HST data have very small formal error bars (typically $\pm$0.02 to 0.03 magnitudes) but somewhat larger systematic errors (judged from the scatter to be of order 0.1 magnitudes) owing to incomplete sampling of the rotational lightcurve (see below).   

We adopt $H$ = 17.96$\pm$0.10 as our best estimate of the mid-light absolute magnitude of the nucleus, indicated in the figure by a dashed line.  Figure (\ref{absmags}) shows that the Stevenson et al.~(2012) (labeled S12) and Novakovic et al.~(2014) (N14) absolute magnitudes  are brighter than $H$ by about 2.0 and 0.6 magnitudes (factors of $\sim$6 and 1.7), respectively, presumably as a result of near-nucleus dust contamination of the photometry.   Even the measurements by Drahus et al.~(2015) (D15), taken  in 2014 (three years after the release of material), suffer  dust contamination of $\sim$0.3 magnitudes ($\sim$30\%).  

The effective scattering cross-section, $C$ [km$^2$], is computed from the photometry using

\begin{equation}
C = \frac{1.4\times10^6}{p_V}10^{-0.4 H}
\label{C}
\end{equation}

\noindent where $p_V$ is the (unmeasured) geometric albedo.   Asteroids orbiting  near 3 AU tend to be C-types with low albedos ($p_V \sim$ 0.05) as opposed to S-types, which have brighter surfaces ($p_V \sim$ 0.2).   In this regard, we note that Stevenson et al.~(2012) measured  a nearly neutral optical color, B-R = 1.03$\pm$0.04 (the color of the Sun is B-R = 0.99$\pm$0.02; Holmberg et al.~2006) as did  Novakovic et al.~(2014) with V-R = 0.40$\pm$0.14 (the solar color is V-R = 0.35$\pm$0.01); both colors are consistent with C spectral type.  However, Figure (\ref{absmags}) shows that both color determinations were made in the presence of strong dust contamination, and so do not directly refer to the  nucleus.  

To proceed, we assume that 331P is a C-type, with $p_V$ = 0.05.  Then, with $H$ = 17.96, Equation (\ref{C}) gives $C$ = 1.8 km$^2$ and the radius of an equal-area circle is $r_n = (C/\pi)^{1/2}$, giving  $r_n$ = 0.76$\pm$0.01  km.    The cross-section would be four times smaller, and the radius two times smaller, if the albedo were instead $p_V$ = 0.2.   Our estimate is small compared to  the upper limit, $r_n \lesssim$ 4.2 km, reported by Stevenson et al.~(2012),  because of the effects of dust contamination in their early-time photometry.

We model the shape of the fading lightcurve in Figure (\ref{absmags}) using the procedure described in Jewitt et al.~(2017).  Briefly, particles released from the nucleus at time $T_0$ are swept out of the photometry aperture on a timescale that depends on the particle size.  Small particles are accelerated strongly by radiation pressure while large particles linger longer, such that the average  radius of particles remaining in the photometry aperture increases even as the total cross-section in the aperture decreases.  For a power-law distribution of particle radii, in which the number of particles with radii between $s$ and $s + ds$ is $n(s)ds = \Gamma s^{-q} ds$ ($\Gamma$ and $q >$ 3 are constants), the  time dependence of the  cross-section may be written (Jewitt et al.~2017)

\begin{equation}
C(t) = C_n + K (t-T_0)^{6-2q}
\label{fade}
\end{equation}

\noindent where $C_n$ is the cross-section of the nucleus, $K$ is a constant and $t \ge T_0$.   Equation (\ref{fade}) is strictly valid provided that the initial particle velocity is small compared to the velocity acquired under the action of radiation pressure.  (The Equation also does not apply when $t - T_0 \lesssim$ 2 months, because ejected fragments have then not yet reached the edge of the photometry aperture.  However, the first observations have $t - T_0 \sim$ 9 months, so this is not a serious limitation).  We set $T_0$ = 2011.5 and made a least squares fit of Equation (\ref{fade}) to the data, finding $C_n$ = 1.81$\pm$0.18 km$^2$ and $q = 4.1\pm0.2$.  The fit is shown as a solid line in Figure (\ref{absmags}).   $C_n$ gives $H$ = 17.98$\pm$0.04 and $r_n  = 0.76\pm$0.02 km for $p_V$ = 0.05, in agreement with the determinations above.  In the following, we use $r_n$ = 0.8$\pm$0.1 km as a working value.  The derived value of $q$ is slightly steeper than $q =$ 3.7 to 4.0 obtained from the Monte Carlo models (Section \ref{trail}) but, given the heterogeneous nature of the photometric data (e.g.~different filters and different apertures were used to compile Figure \ref{absmags}), and the different assumptions made in each method, the difference is  not significant.

All HST observations show systematic, short-term brightness variability indicative of rotation with a two-peaked period, $P$, near 3 hours.  The temporal sampling of HST (images are acquired within windows of about 40 minutes separated by gaps of about an hour) prevents continuous phase coverage of the lightcurve in any given month. Observations from different years (Table \ref{geometry}) are, for the most part,  too widely separated to be unambiguously phased together.  The exception is that data from UT 2015 December 25 and 28 (3 day separation, corresponding to about 22 rotations) are close enough together that we can meaningfully link the data to derive the rotation period, albeit with aliases separated by about 0.014 day$^{-1}$.  Phase dispersion minimization applied to the absolute magnitudes from these two dates gives  comparably acceptable, two-peaked periods $P$ = 3.170$\pm$0.010 hour and $P$ = 3.259$\pm$0.010 hour. The latter of these is close to the period deduced from ground-based data, namely $P$ = 3.24$\pm$0.01 hour (Drahus et al.~2015).     In Figure (\ref{primary}) we show the data from UT 2015 December 25 and 28 phased to  $P$ = 3.259 hour (yellow circles) compared to the lightcurve from Drahus et al.,  rescaled to the same period and shifted in $H$ using V-R = 0.4 to account for the use of different filters. The phase offset between data from 2014 and 2015 is unknown.  Figure (\ref{primary}) shows that the basic features of the lightcurve are well reproduced, although with small differences.  For example,  the lightcurve range, $\Delta V = 0.25$ magnitude, is $\sim$0.05 magnitudes larger than in the ground-based data.  This likely reflects the combined effects of dilution of the nucleus photometry in 2014 owing to near nucleus dust that had cleared by 2015 (Figure \ref{absmags}), and a year-to-year change in the viewing geometry with respect to the projected spin-pole of 331P. A consistent but less precise period,  $P = 3.25\pm0.05$ hours, was determined from the HST image sequence on UT 2017 February 13.

Approximating the shape of 331P by a prolate ellipsoid with semi-axes $a\times b \times c$, with $a = b$, and $c/a \ge 1$, in rotation about a short axis, we estimate the axis ratio from the lightcurve using $c/a = 10^{0.4\Delta V}$.  Substituting $\Delta V$ = 0.25 magnitudes gives  $c/a \sim$ 1.25.  The gravitational acceleration acting on the tip of the nucleus ellipsoid exceeds centripetal acceleration provided the density exceeds a critical value

\begin{equation}
\rho_c = \frac{3\pi}{G P^2} \left(\frac{c}{a}\right)^2.
\label{rho_c}
\end{equation}

\noindent Substituting $P$ = 3.26 hours gives $\rho_c$ = 1600 kg m$^{-3}$, which is  of firm significance only if  331P can be meaningfully represented as strengthless.   For comparison, the mean densities of  C-type ($\rho \sim 1500\pm500$ kg m$^{-3}$) are compatible with this value within the uncertainties (Hanus et al.~2017).    With $\rho$ = 1600 kg m$^{-3}$ and $r_n$ = 0.8 km, the nucleus mass is $M_{n} =  3\times10^{12}$ kg.  The gravitational escape velocity from a non-rotating sphere with  density $\rho$ = 1600 kg m$^{-3}$ and radius $r_n$ = 0.8 km is $V_e$ = 0.8 m s$^{-1}$ but rotation may reduce the effective escape speed to near zero at the tips of the long    axis.

\subsection{Debris Trail}
\label{trail}

The trail is the dominant morphological feature of 331P.  Figure (\ref{PA}) shows the position angle of the trail, $\theta_{PA}$, measured with respect to the nucleus and plotted as a function of the true anomaly, $\nu$.  The trail angle and its uncertainty (typically  $\pm$0.1\degr) were measured using  determinations of the location of peak brightness at different projected distances from the primary nucleus.  The position angle of the negative heliocentric velocity vector, $\theta_{-V}$, obtained from JPL Horizons (\url{https://ssd.jpl.nasa.gov/horizons.cgi}), is also shown in the figure as a solid black line.  There is generally excellent agreement between $\theta_{PA}$ and $\theta_{-V}$.   On the other hand, there is no correspondence between $\theta_{PA}$ and the projected antisolar vector, $\theta_{-\odot}$, which undergoes much more dramatic excursions in the range 65\degr~$\le \theta_{-\odot} \le$ 169\degr~during   the period covered by the observations (see Table \ref{geometry}).  This confirms that the basic  morphology of the trail reflects the distribution of  large particles confined to the orbit, as  inferred by Moreno et al.~(2012) and Stevenson et al.~(2012).  

The dominance of large particles is consistent with the action of radiation pressure, which truncates the size distribution  by quickly sweeping small particles away.  The dispersal timescale  is

\begin{equation}
\tau_{rad} \sim \left(\frac{2 \ell_T}{\beta g_{\odot}(r_H)}\right)^{1/2}
\label{trad}
\end{equation}

\noindent where $\ell_T$ is the length of the trail under consideration, $g_{\odot}(r_H)$ is the gravitational attraction to the Sun at distance $r_H$, and $\beta$ is the dimensionless radiation pressure factor, defined such that $\beta g_{\odot}(r_H)$ is the particle acceleration.  With a nearly circular orbit at 3 AU, we write $g_{\odot}(3) = 7\times10^{-4}$ m s$^{-2}$.  Parameter $\beta$ is a function of the wavelength, and of many unknown particle parameters, including the composition, shape, size and porosity.  For spherical, uniform dielectric particles we take $\beta \sim 10^{-6}/s$, where $s$ is the particle radius in meters (Bohren and Huffman 1983).   To give an example, the trail in 2015 December extends beyond the 160\arcsec~field of view of WFC3, corresponding to $\ell_T \sim 2.4\times10^8$ m, in the plane of the sky, given the geocentric distance  $\Delta$ = 2 AU.  Substituting into Equation (\ref{trad}) gives $\tau_{rad} \sim 28 s^{1/2}$, with $\tau_{rad}$ in years.  Setting $\tau_{rad}$ = 4.5 years (the time elapsed between 2015 December and the ejection in 2011 July), gives $s \sim$ 2.5 cm, meaning that  smaller particles have been swept away.

Particles ejected from the nucleus  increase their distance from  the orbit plane in proportion to $t$, the time of flight, but are accelerated down the trail by a distance proportional to $t^2$, as a result of the steady acceleration by solar radiation pressure.   Consequently, the trail width perpendicular to the orbit plane, $W_{\perp}$, is expected to increase with distance from the nucleus, $\ell_T$, as $W_{\perp} \propto \ell_T^{1/2}$.

We used the composite image from UT 2018 May 17  to test this expectation by measuring the width of the dust trail as a function of distance along the trail.  Observations on this date were taken from the smallest angle (-1.5\degr) between HST and the orbit plane (Table \ref{geometry}) and thus suffer least from the effects of projection.       The results are plotted in Figure (\ref{sb_perp}).   The widths were determined by visual inspection from the FWHM of the perpendicular surface brightness profile averaged over segments  whose widths are indicated by horizontal bars in the figure. The expected $W_T \propto \ell_T^{1/2}$ dependence provides an acceptable fit to the data, shown by the red line least-squares fit in the figure.   

The velocity vs width relation is (Jewitt et al.~2014a)

\begin{equation}
V_{\perp} = \left(\frac{\beta g_{\odot}}{8 \ell_T} \right)^{1/2} W_{\perp}
\label{speed}
\end{equation}

\noindent in which $V_{\perp}$ is the ejection velocity perpendicular to the orbit plane.    We substitute $\beta = 10^{-4}$ for centimeter-sized particles, $g_{\odot} = 7\times10^{-4}$ m s$^{-2}$ for the solar gravity at 3 AU, $W_{\perp}$ = 1700 km and  $\ell_T = 10^5$ km (Figure \ref{sb_perp}) to find $V_{\perp} \sim$ 2 cm s$^{-1}$.  The result is approximate for two reasons; first, because the observations  were taken  -1.5\degr~from the orbit plane and thus suffer from projection and second, because Equation (\ref{speed}) is strictly valid  only when the particle speed induced by radiation pressure is large compared to the ejection speed. Nevertheless, the trail width measurements in Figure (\ref{sb_perp}) are consistent with an extremely small perpendicular velocity.

Figure (\ref{sb}) shows the surface brightness (black line) as a function of angular distance from the primary nucleus along the axis of the trail using data averaged from the independent profiles from UT 2015 December 25 and 28  to improve the SNR.  The surface brightness was computed within a region extending $\pm$1.2\arcsec~from the axis, with sky background determined from the average of adjacent strips of equal width centered 2.4\arcsec~above and below the trail.   The profile is  binned to 0.4\arcsec~resolution along the trail direction in order to improve the signal-to-noise ratio.   Numerous trailed field stars and galaxies contaminate the profiles, as do scattered light flares from bright objects near the HST field and artifacts from imperfect flattening.  The most egregious of these are marked in Figure (\ref{sb}) by asterisks, $*$, and the region from 58\arcsec~to 77\arcsec~is omitted entirely owing to background light that could not be adequately removed.  The axial profile in the UT 2015 December data clearly peaks about 40\arcsec~(58,000 km in the plane of the sky) behind the primary nucleus.

We used a Monte Carlo model (Ishiguro et al.~2007) to reproduce the measured surface brightness profile in Figure (\ref{sb}).   The model considers the motions of ejected particles under the action of radiation pressure and solar gravity.  While such a multi-parameter model cannot produce unique solutions, this approach does allow us to limit the possible range of several important particle parameters.  For example, in gas drag acceleration, the initial particle speed varies with the particle radius, $s$, as $V \propto s^{-1/2}$ (Whipple 1950).  We find that no models with this velocity - size dependence produce a convincing match to the data because the trail widths from  faster particles exceed the measured trail width, confirming a result from Moreno et al.~(2012).    Models with  ejection on UT 2011 July 1, low size-independent velocity $V$ = 5 to 10 cm s$^{-1}$, and a power-law size distribution with index $q$ = 3.7, give a reasonable match to the data up to $\sim$20\arcsec~west of the nucleus but, at larger angles, the measured surface brightness is lower than in the $q$ = 3.7 model (Figure \ref{sb}).  This means that the largest particles, which have traveled the smallest distance from the nucleus, are adequately represented by $q$ = 3.7 but smaller particles,  accelerated by radiation pressure to larger distances, are not.  We could find no single power law distribution to improve the fit far west of the nucleus without losing the fit nearer the nucleus.   Therefore, we fitted a broken power law to the surface brightness profile (Figure \ref{sb}).  We obtained an improved fit with a break at $\beta_{crit} = 10^{-5}$ (corresponding to $s \sim$ 10 cm  particles) and $q$ = 4.0 for $\beta \le \beta_{crit}$ ($s \ge$ 10 cm) and $q$ = 3.0 (blue line in Figure \ref{sb}) or even 2.5 (green line in the figure) for larger $\beta$ ($a <$ 10 cm).  The physical significance of these fits is, of course, questionable, since by adding a broken power law we are adding more free parameters to a multi-parameter model.  However, the results do  broadly support the work by Moreno et al.~(2012),  but using independent data that samples the debris four years after launch.  The results are also broadly compatible with the index $q = 4.1\pm0.2$ deduced from the fit to fading lightcurve (Section \ref{primary} and Equation \ref{fade}).

In order to measure the   integrated brightness of the trail we used a 3.2\arcsec$\times$48.0\arcsec~rectangular section  with the long axis extending along the trail  to the west.  More distant portions of the trail are  very faint and introduce excessive uncertainty owing to variations in the background.  As a measure of the background, we determined the median signal in  two same-sized flanking strips above and below the trail and subtracted their average.  Measurements on UT 2015 December 25 and 28 are close enough together in time that the trail is unlikely to have changed, and so permit a measure of the effect of interfering background sources and scattered light in the data.   The integrated trail magnitudes were  V = 22.24 and 22.18 on December 25 and 28, respectively, indicating agreement at the $\sim$6\% level.  The average of these measurements, together with Equations \ref{HV} and \ref{C}, gives a dust scattering cross-section $C = 27\pm3$ km$^2$, again assuming $p_V$ = 0.05.  This  trail cross-section  is spread over the $\sim3.3\times10^8$ km$^2$ projected area of the measuring rectangle, corresponding to an average optical depth of only $\sim6\times10^{-9}$.  This value lies within the range of optical depths reported for the large-particle trails of short-period comets (Ishiguro et al.~2009).

Given that the trail is optically thin, its mass, $M$, can be written  $M \sim \rho \overline{a} C$, where $\overline{a}$ is the cross-section weighted mean particle size.  The trail rectangle length, $\ell_T$, is $\ell_T \sim 7\times10^4$ km, and the observations were taken $\sim$4.5 years after the particle release, giving $\tau_{rad} = 1.4\times10^8$ s.  Substitution into Equation (\ref{trad}) gives $\beta \sim 10^{-5}$,  corresponding to a minimum particle size in the measured portion of the trail of $\sim$10 cm.    Strictly, $\ell_T$ is a lower limit to the visible length of the trail because of the effects of projection, but we can confidently conclude that particles smaller than centimeters in size will have already left the trail. With $\overline{a}$ = 0.1 m, we obtain a lower bound to the trail mass, $M \gtrsim 2\times10^8$ kg.  For comparison, Moreno et al.~(2012) found $M \gtrsim5\times10^8$ kg while Stevenson et al.~(2012) found $M \gtrsim5\times10^7$ kg, both in data from 2012.  The ratio of the mass in the optically dominant particles relative to the nucleus mass is $M/M_{n} \sim 10^{-4}$.

\subsection{Fragments}
\label{fragments}
\begin{figure}
\epsscale{0.99}
%\plotone{SB.pdf}
%\plotone{images.pdf}
%\plotone{keycode2.pdf}
\plotone{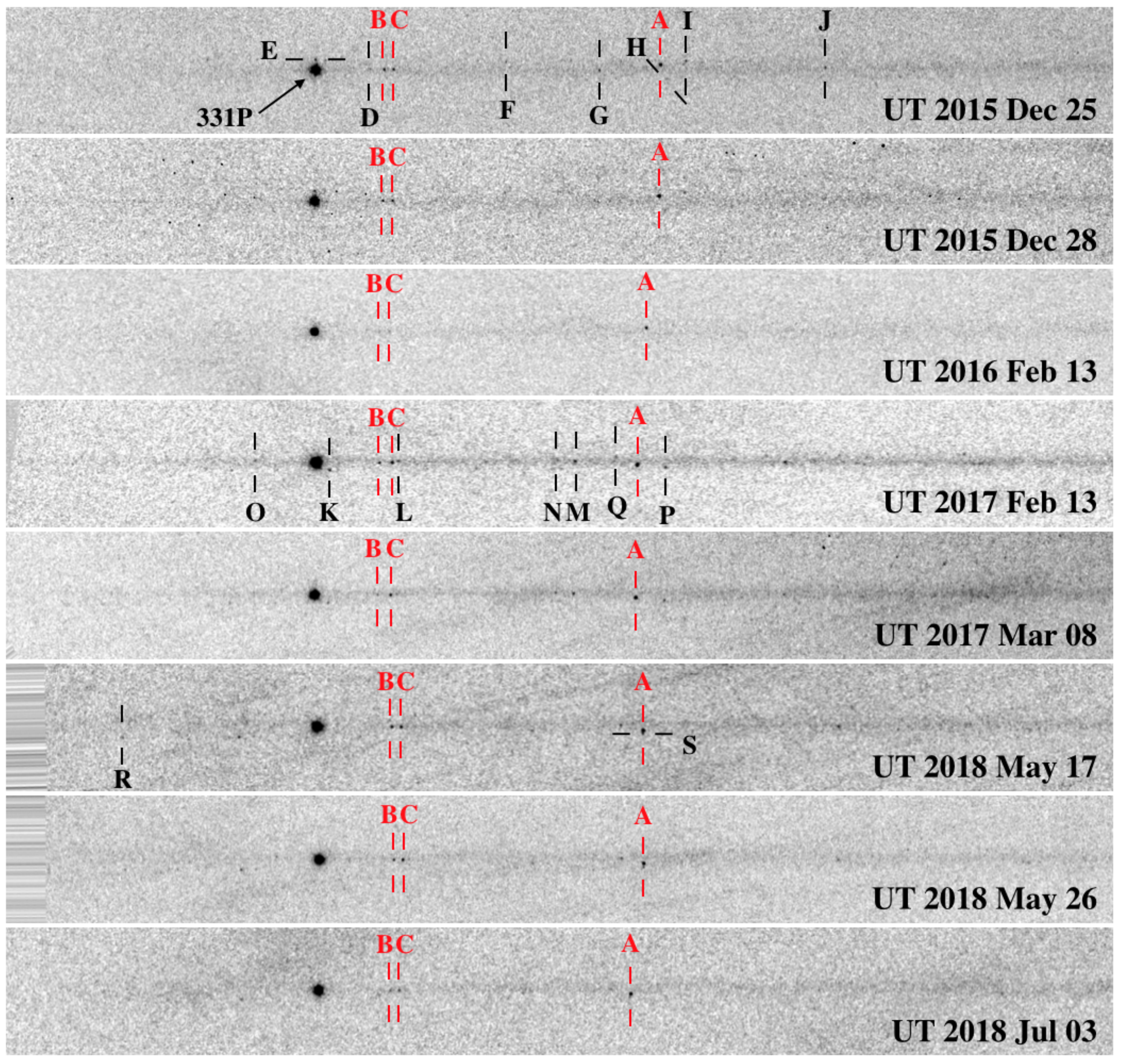}

\caption{Composite of images scaled to $\Delta$ = 2.011 AU such that the width of each panel corresponds to 92,000 km at the distance of 331P.  Fragments 331P-A, -B and -C are labeled in red in each panel.  Other fragments are identified only where they first appear.  Diagonal streaks, especially noticable in the UT 2018 May 17 panel, result from incompletely removed trailed background objects.  There is no correction for the changing phase angle.  \label{keycode}}
\end{figure}

\begin{figure}
\epsscale{0.8}
%\plotone{SB.pdf}
%\plotone{images.pdf}
%\plotone{absmags3.pdf}
\plotone{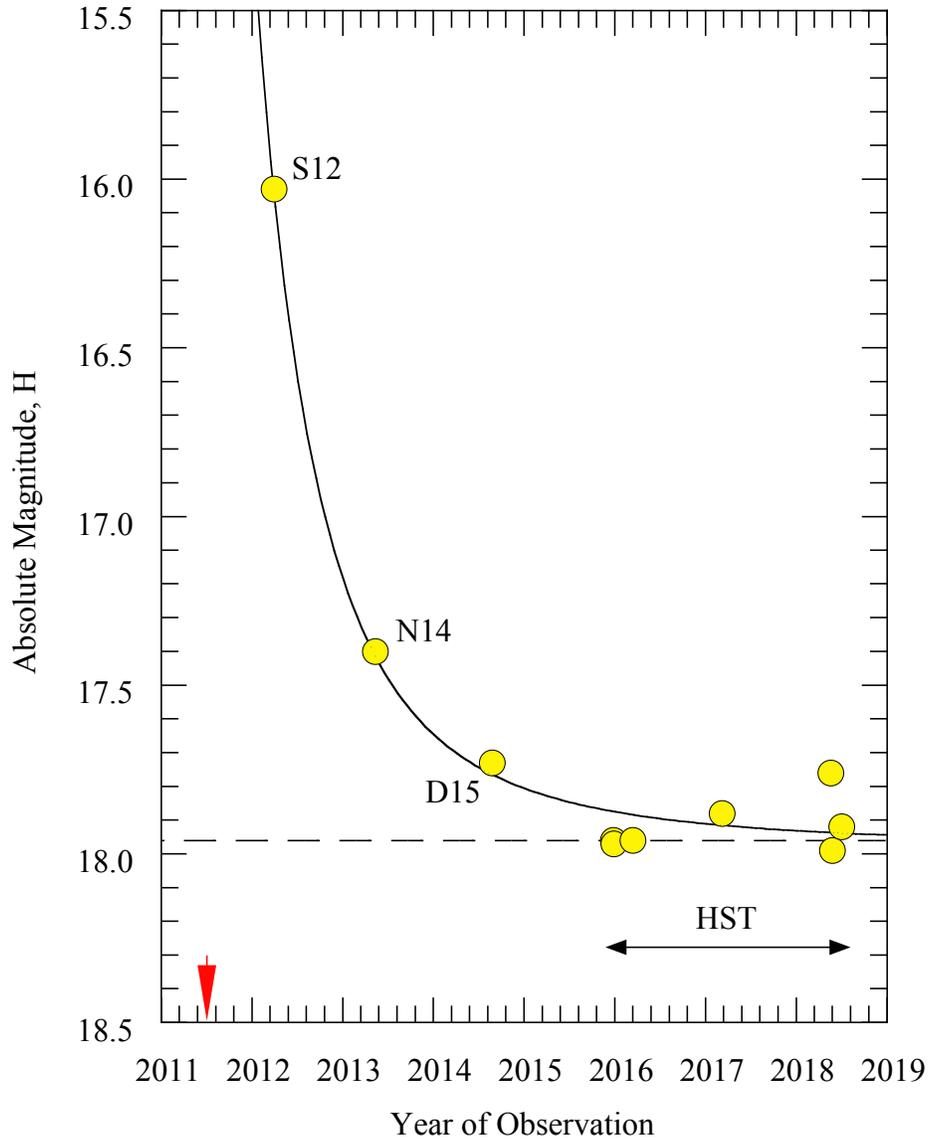}

\caption{Absolute magnitude vs.~date of observation for primary nucleus 331P (yellow circles). Letters indicate measurements from the literature (S12 = Stevenson et al.~(2012), N14 = Novacovic et al.~(2014), D15 = Drahus et al.~(2015)) color-corrected to the V filter.  The downward pointing red arrow marks 2011 July 1, the nominal date of initiation of the activity in 331P, while the dashed horizontal line shows the absolute magnitude of the primary nucleus in the absence of dust. The black line is the best fit of the fading Equation (\ref{fade}).   \label{absmags}}
\end{figure}

 The visibility of the fragments varies from panel to panel in Figure (\ref{keycode})  both because  different numbers of images and total on-source integration time were acquired in different orbits and because of the varying geometry of observation (Table \ref{geometry}).  For example, few fragments can be detected on UT 2016 February 13 and July 03 because of the larger geocentric distances and phase angles on these dates, causing a geometrical dimming by $\sim$0.8 magnitude relative to the observations from UT 2015 December 25.  The composites from UT 2015 December 28 and 2018 May 17 are diminished by the smaller number of images secured on these dates (Table \ref{geometry}).   
 
Most of the fragments are detected in the first and deepest composite image (from UT 2015 December 25) which benefits from both a large number of images and a favorable geometry.  331P-A, -B and -C, as well as 331P itself, are unambiguous in all panels and are marked in red in each panel of Figure (\ref{keycode}) to act as guideposts.  Other fragments are marked when first noticed but are not labeled in subsequent panels to avoid excessive clutter.  For example, Fragment Q is first noticed and labeled in UT 2017 February 13 and is visible again (but not labeled) in data from UT 2018 May 17 and May 26 (and, possibly, UT 2018 July 03). Fragment S is slightly below Fragment A in Figure (\ref{keycode}) when first noticed (UT 2018 May 17) and is visible again in UT 2018 May 26 and July 03.  Most of the fragments first detected after 2015 are close to the detection threshold and could have been missed in UT 2015 December 25 as a result of temporal brightness changes (due to rotation?), and in other visits because of the changing limiting magnitude.   331P-R in UT 2018 May 17, for example, has no convincing counterpart in earlier data, although it is faintly visible again on May 26.   Overall, the fragment trail is challenging to study because it is spatially complex and temporally under-sampled (e.g.~the long interval between the high quality images from UT 2015 December 25 and 2017 March 08).  Firm linkages across the dataset await a detailed dynamical study to be undertaken separately.  

%Mention something about simultaneous vs staggered fragment ejection
 
In Table (\ref{photometry}) we list the mean absolute magnitudes of the fragments using data from all eight visits.  To minimize the effects of background light and, especially, the interfering effects of imperfectly removed cosmic rays, we present only photometry within a 5 pixel (0.2\arcsec) radius circular aperture, with sky subtraction from a contiguous annulus extending to 105 pixels (4.2\arcsec).  The table lists the mean, $H_5$, and median, $H_m$, magnitudes showing that, in most cases the two estimates of the central value are in good agreement.  The uncertainty on $H_5$ is just the standard error on the mean from different orbits except that, for fragments measured in only one orbital visit, we have set the uncertainty to 0.1 magnitudes.  The last column of Table (\ref{photometry}) lists the fragment radius, computed from $H_m$ and Equation (\ref{C}).  The uncertainties on $r_n$ are formal only, and underestimate the true uncertainties because of the unknown albedos and, to a lesser extent, phase functions of the fragments.

We first consider the HST photometry for 331P, 331P-A, -B and -C, the four objects that were measured in all seven visits.  The measurements shown in Figure (\ref{4objects}) are  vertically offset for clarity.  Lines in the figure show least-squares fits to the photometry as a function of time and show that in no case is there evidence for secular fading or brightening of the fragments across the full 2.5 years of the HST data set.  This is strong evidence that the fragments are inert (asteroid-like) bodies, not  releasing dust particles.  The lack of comet-like activity is further suggested by  the absence of comae or tails on the fragments, all of which are unresolved at HST resolution (0.08\arcsec~FWHM).

Other than 331P itself, only 331P-A is bright enough to permit the determination of a meaningful rotational lightcurve.  We compile photometric observations using a 0.2\arcsec~radius aperture from UT 2015 December 25 and 28 and UT 2017 February 13 in Figure \ref{secondary}. The three visits are so widely spaced (and therefore so seriously aliased) that we cannot accurately phase the three epochs together.  Instead, we have arbitrarily shifted the time axes to match that of the observations from UT 2015 December 25.  The lightcurve shows a large and stable range of $\Delta V \sim$ 1.3 to 1.5 magnitudes with a single maximum and two clear minima separated by about  4.5 hours.    If the lightcurve is due to rotation (i.e.~is two-peaked), Figure \ref{secondary} suggests a period near 9 hours.    The large photometric range and the V-shaped brightness minimum (Lacerda and Jewitt~2007) are consistent with the lightcurve of a contact binary, as noted by Drahus and Waniak~(2016).  If this is the case, we represent the body by two  ellipsoidal components, each with semi-axes $a \times a \times c$ and $c \ge a$, touching nose-to-nose.  The maximum and minimum projected cross-sections, for a rotation axis in the plane of the sky, would then be 2$\pi ac$ and $\pi a^2$, respectively, giving axis ratio 

\begin{equation}
\frac{c}{a} = \frac{10^{0.4\Delta V}}{2}.
\end{equation}

\noindent  Substituting $\Delta V$ = 1.5 magnitudes gives $c/a$ = 2.0.  Assuming $p_V$ = 0.05 and $H_m$ = 22.25 (Table \ref{photometry}), we estimate $a$ = 65 m, $c$ = 130 m such that the overall structure would resemble a peanut four times longer than any perpendicular dimension, rotating in about 9 hours.  The net gravitational attraction to the center would exceed the centripetal acceleration provided the density exceeds a critical value, 

\begin{equation}
\rho_c = \frac{12\pi}{G P^2} \left(\frac{c}{a}\right)^2,
\end{equation}

\noindent where $G = 6.67\times10^{-11}$ N kg$^{-2}$ m$^2$ is the gravitational constant.  Substituting $c/a$ = 2, $P$ = 9 hours, we find $\rho_c$ = 2150 kg m$^{-3}$.  This value is  slightly bigger than the limiting  value for 331P (1600 kg m$^{-3}$; Equation \ref{rho_c}), and than the mean density of C-types ($\rho \sim 1500\pm500$ kg m$^{-3}$) but comparable to the mean density of S-type asteroids ($\rho \sim 2500\pm500$ kg m$^{-3}$) according to Hanus et al.~2017).  However, the partial lightcurve (Figure \ref{secondary}) and the approximate nature of this density estimate are not sufficient to prove the contact binary hypothesis.   Most importantly, even very small values of the material cohesion can negate simplistic density estimates like those above, based on its assumed absence (e.g.~Holsapple and Housen 2019, Hirabayashi and Scheeres 2019). 
%%%%

The contact binary interpretation cannot be regarded as unique.  For example, one possible alternative explanation for the 1.3 to 1.5 magnitude lightcurve range is that  331P-A is a monolithic body with an axis ratio $c/a \sim$ 3 to 4.   Another possibility is that 331P-A is a sphere supporting a hemispheric azimuthal albedo which varies by a factor of 3 to 4.   We simply remark that comparably extreme bodies are  rare or unknown in the asteroid population.  For example, only 3 out of 4986  asteroid  lightcurves in the compilation by Warner et al.~(2019) have $\Delta V \ge$ 1.4 magnitudes ($c/a \gtrsim $ 3.6:1) and these might also be elongated  contact binaries.    Except for the pathological case of Saturn's satellite Iapetus (where synchronous rotation and an external, anisotropic source of dust are responsible) 4:1 hemispheric albedo variations are unknown in the solar system.  Therefore, while   alternative explanations cannot be formally excluded, we regard them as less convincing than the contact binary model.   Such an object would presumably form by mutual attraction of ejected fragments, as occurred in the gently disrupted P/Shoemaker-Levy 9 (Movshovitz et al.~2012) and consistent with numerical simulations showing complex and drawn-out interactions amongst the particles ejected from rotationally unstable bodies (Walsh et al.~2012, Boldrin et al.~2016).  Some numerical experiments suggest that asteroid contact binaries can be  produced in abundance by  gravitational re-accumulation of fragments following asteroid disruption (Campo Bagatin et al.~2020), as may be the case for 331P-A.
 %%%%

The  fragment size distribution is shown  as a cumulative distribution  in Figure (\ref{size_plot}), with error bars in the latter figure  assigned according to Poisson statistics, and also schematically in Figure (\ref{schematic}).  Flattening of the  distribution  at $r_n \lesssim$ 0.05 km reflects observational incompleteness in the detection of small (faint) fragments.   Lines of slope $q$ = 3, 4 and 5 are shown for reference.  While noisy, the data are  consistent with $q \sim 4$ to 4.5, slightly steeper than $q = 3.7\pm0.1$ to $q$ = 4.0 as inferred from Monte-Carlo models of large particles in the trail (Moreno et al.~2012, Section \ref{trail}) and   $q = 4.1\pm0.2$ found from the fading curve in Figure (\ref{absmags}).  If the densities of the fragments are the same as the density of 331P, we can use the radii in Table (\ref{photometry}) to estimate the fractional mass in fragments as $M/M_{n} \sim$ 0.01.  This is 100$\times$ larger than the mass in diffuse material (Section \ref{trail}), consistent with a mass distribution dominated by the largest bodies, even across the full cm to 100 m size range.   We note that Moreno et al.~(2012) inferred a much larger $M/M_n \sim$ 0.2, in part because they inferred a 331P nucleus radius of only $\sim$100 to 150 m,  which our nucleus photometry  rules out.

 In this regard, it is interesting to compare the fragments of 331P with the carefully measured boulder size distributions of other asteroids, determined from in-situ data.   Boulders on the 0.5 km diameter, C-type asteroid 101955 Bennu follow a $q$ = 3.9$\pm$0.3 differential power law (Dellagiustina et al.~2018) similar to $q \sim$ 4 to 4.5 in 331P.  Measurements of gravity data from Bennu indicate that the  surface boulders accurately reflect the size distribution in the deep interior, and show that the largest ``particle'' inside the body of Bennu is $\sim$ 150 m in size  (Tricarico et al.~2021).  This is comparable to the $\sim$100 m equivalent spherical size of 331P-A.  Given these similarities, we speculate that the material ejected from 331P is just the fragmentary matter from near the surface of 331P, destabilized and launched by rotation.  
 
The surface density of large boulders on Bennu is unpublished but on similar object Ryugu, the  density for boulders having radii $>$50 m is $\Sigma \sim$0.5 to 1 km$^{-2}$ (Michikami and Hagermann~2021).  Scaling to the surface area of  331P, $A \sim 4\pi r_n^2 \sim 8$ km$^2$, we would expect $\Sigma A \sim$4 to 8 objects of 50 m scale, when we nominally   detect 15 objects with radii $>$50 m in 331P (Table \ref{photometry}).  The approximate  agreement is consistent with the idea that most 331P fragments were pre-existing bodies (boulders) located on or near the surface of 331P and liberated from it by a rotational instability.

\clearpage

\section{DISCUSSION}

\subsection{Origin Scenario}

We envision the following steps in the formation of 331P and its associated fragments and debris trail.

\textbf{Formation:}  331P was  liberated from its parent body (20674 (1999 VT1)) by impact only $\tau_C \sim$1.5 Myr ago, as indicated by its membership in a tightly clustered group of nine asteroids (Novakovi{\'c} et al.~2014).  331P (radius $\sim$0.8 km)  is the smallest of the known members of this group but many smaller members presumably remain to be discovered (and, when they are, should be examined for active asteroid status, since their spin-up timescales will be shorter than that of 331P).  The precursor body  was the target of a cratering collision, not a full blown disruption,  and most of the parent body mass remains in the  9 km radius cluster member 20674 (1999 VT1).  

\textbf{Spin-Up:} Small size renders 331P susceptible to spin-up by radiation (YORP) torques on timescales $\tau_Y \sim \tau_C$.  Specifically, a fit to the measured   spin-up timescales of asteroids as a function of radius gives  $\tau_Y \sim 4 r_n^2 (r_H/3)^2$, where radius $r_n$ is expressed in kilometers, $r_H$ is the semimajor axis in AU and $\tau_Y$ is in Myr, with a wide dispersion due to shape and thermal differences amongst the asteroids (Jewitt et al.~2015).  Substituting $r_n$ = 0.8 km and $r_H$ = 3 AU for 331P gives $\tau_Y \sim$ 2 Myr, equal to $\tau_C$ within the considerable uncertainties on both timescales.  If interior ice was exposed by the collisional ejection of 331P, then sublimation torques could also accelerate the spin on a short timescale. %Damping of the spin-state is negligible on timescale $\tau$ (e.g.~the dissipation timescale is $\sim2\times10^8$ years using the canonical formula from Burns et al.~1973).   %This raises the question of why the lightcurve of 331P appears consistent with principal axis (minimum energy) rotation.  Future observations could reasonably be targeted to search for lightcurve evidence of non-principal axis rotation.

\textbf{Rotational Instability:} Rapid rotation (Figure \ref{primary},  Drahus et al.~2015) and the observed independence of the ejection velocity on particle size (Section \ref{fragments}, also Moreno et al.~2012) are smoking guns for the role of rotational instability.  (Gas drag gives a velocity vs.~size relation, $V \propto s^{-1/2}$, which is incompatible with the data).   Additionally, the observation that the ejection velocities ($V \sim$ 5  to 10 cm s$^{-1}$) are small compared to the gravitational escape speed, $V_e \sim$ 80 cm s$^{-1}$, is naturally explained by rotational instability because, in this scenario, equatorial material is effectively already travelling at the escape speed and only a very tiny velocity increment is needed to induce escape.   

The break-up of an aggregated body involves complex and incompletely understood interactions between escaping fragments (Walsh et al.~2012, Davis and Scheeres 2020, Hu et al.~2021).  For example, material leaving radially at 5 cm s$^{-1}$ takes $\sim$4.3 hours to travel one nucleus radius, during which time the underlying source body has completed a full rotation, giving time for  torques from the aspherical primary to deflect the material.  In numerical simulations, interactions between escaping particles lead to energy dissipation and can result in trapping in-orbit and even satellite  formation (Walsh et al.~2012, Hu et al.~2021).  As a result of these complex interactions, the estimated dynamical timescales in rotationally fissioned bodies are model-dependent and, potentially, very long (e.g.~days to years; Boldrin et al.~2016).    Some models show the formation of contact binaries in the ejecta from low energy collisions (Campo Bagatin et al.~2020), an intriguing result in view of the evidence that 331P-A is a contact binary.   For these reasons, it is presently difficult to relate the observations of 331P to specific models which offer, at best, only statistical predictions of the outcome.

%We note that it is possible that the fragments of 331P were ejected in a staggered fashion, with the more distant fragments being ejected earlier.  Staggered ejection was found in the disrupted body P/2013 R3 where, for example, down-trail Fragment D was probably the first to leave (Jewitt et al.~2017).  In view of the small range of measured motions (Figure \ref{keycode}) and the substantial influence of projection, however, we leave an examination of this possibility to a future paper.

The numbers of fragments projected east and west of 331P are not comparable, showing that the ejection was not symmetric.   We find 2 or 3 fragments to the east of 331P and 16 or 17 to the west, rather than a more nearly 1:1 ratio (Figure \ref{keycode}).  This asymmetry  suggests the approximately simultaneous release of the fragments from a localized region of failure on the surface of 331P, perhaps corresponding to one of the unstable  tips of the rotating ellipsoid representing the shape of the body.  Alternatively, it is possible that a single ejected body  partially disintegrated  into co-moving  fragments  during the ejection.  An analog might be the 160 m boulder ``Otohime Saxum'' on Ryugu, whose fractured appearance (see Figure A5 in Yokota et al.~2021) suggests that it would break up if it were to be disturbed.  Still another possibility is that the fragments initially exposed buried ice, leading to sublimation on their sunward sides and a systematic non-gravitational acceleration away from the Sun, as is commonly observed in the fragments of split comets (e.g.~Boehnhardt 2004).  We regard this as not very likely because non-gravitational acceleration varies as $1/r_n$, and the fragments of 331P are large compared to those in typical split comets. Also, at 3 AU, the rate of sublimation of water ice and the resulting recoil force are $\sim$16 times smaller than at 1 AU, where split comets are most commonly observed.  Lastly, the HST images provide no evidence for dust release from any fragment.  However, we need detailed dynamical studies to more firmly assess the possibility of non-gravitational acceleration as the cause of the east vs.~west asymmetry.

\textbf{Dispersal:} The overall dispersal time of the trail particles is set by radiation pressure and, for the larger bodies that are insensitive to radiation pressure, by the initial velocity.  As noted above, the dispersal timescales for particles (Equation \ref{trad}) are such that sub-centimeter particles have already been swept from the observed portion of the trail.  By comparison, a large fragment launched at $V = 0.1$ m s$^{-1}$,  would take a nominal $2\pi r_H/V \sim 10^6$ years to circulate around the orbit, comparable to the age of the asteroid cluster of which 331P is a member. Eventually, the ejected fragments will dissolve into the background population of outer belt asteroids, but not any time soon.

Conceivably, the observed fragmentation is only the most recent of numerous such events in which mass and angular momentum are dumped, with 331P held at the edge of rotational instability by the continued action of radiation torques.   Given a fractional mass loss $M/M_{n} \sim$ 1\%, 331P could sustain another $\sim$100 similar events.

\subsection{Related Objects}

At least three other active asteroids, P/2013 R3, 311P and Gault, are likely products of rotational instability (Figure \ref{compared}).  The basic properties of these objects are compared with 331P in Table (\ref{breakups}) and briefly described here.  P/2013 R3 showed a fragmented appearance with a dozen or more components   enveloped in a debris trail which had a peak cross-section $\sim$30 km$^2$ (Jewitt et al.~2014b, 2018).   Unlike 331P, P/2013 R3 lacked a dominant nucleus  and consisted instead of a set of closely-spaced ($\sim10^4$ km), debris-enshrouded  bodies each $\lesssim$100 m in size, with a velocity dispersion amongst fragments, $\Delta V \sim$ 30 cm s$^{-1}$.  The fragments continued to evolve with  separation times staggered over $\sim$5 months.  The parent body  of P/2013 R3 was estimated to be $r_n \lesssim$ 0.4 km in radius, therefore containing $\lesssim$15\% of the mass of 331P (equal densities assumed). Morphological differences from 331P include the lack of a dominant nucleus and the progressive disintegration of each observed fragment into smaller bodies.  The relative dustiness of P/2013 R3 results from it being observed with HST sooner after the initial break-up  (two to three months; Figure 12 of Jewitt et al.~2017)  than was 331P (4.5 years) giving less time for radiation pressure sweeping to remove the sub-centimeter debris.  The lack of a dominant nucleus suggests a more severe rotational instability than in 331P, in which the entire body of the initial nucleus was consumed (Jewitt et al.~2014b, Hirabayashi et al.~2014).

Active asteroids 311P/PANSTARRS (2013 P5) (Jewitt et al.~2013, 2018) and 6478 Gault (1988 JC1) (Jewitt et al.~2019, Kleyna et al.~2019) are two other bodies likely experiencing rotational instability.  Both showed multiple tail structures indicating episodic mass loss, in the case of 311P over a nine-month period (Jewitt et al.~2018) and in the case of 6478 Gault spread over at least six years (Chandler et al.~2019). Rotational instability in 6478 Gault is implicated by its short rotational period, 2.55$\pm$0.10 hours (Luu et al.~2021) while the rotational lighcurve of 311P has not been measured  (Hainaut et al.~2014, Jewitt et al.~2017).  As judged by the fractional mass loss estimates (Table \ref{breakups}), 311P and Gault experienced less severe mass loss events  than 331P which, in turn, was less severe than  P/2013 R3.  Mass loss from 311P,  6478 Gault and 331P may be more akin to  avalanching (``surface shedding''; Hirabayashi et al.~2015, Sanchez and Scheeres  2018) than to solid body break-up seen in P/2013 R3.  

Available data thus  expose the rich diversity of outcomes resulting from the rotational instability of asteroids, revealing this to be an exciting subject and a promising area of investigation for planetary astronomers.  The long-lived and slowly spreading fragment chain makes 331P  a particularly attractive target for future spacecraft investigation. Even decades from now, the modest separations between components would allow a single camera-equipped spacecraft to first examine the excited rotational state of the parent body, then investigate the geology of the detachment scar, and then travel down the trail to each of the fragments in succession, including the putative contact binary 331P-A, all the while sampling the boulder distribution in the trail.  Such a mission would provide unprecedented insight into the physics of asteroid rotational instability and fragment dispersal.

\clearpage

\section{SUMMARY}
We study  active asteroid 331P/Gibbs using archival Hubble Space Telescope  data, finding that:

\begin{enumerate}

\item The main component of 331P has a radius $r_n$ = 0.8$\pm$0.1 km (albedo 0.05 assumed), a rotation period $P = 3.26\pm0.01$ hour and a lightcurve range 0.25 magnitudes, indicating an axis ratio $\ge$ 1.25:1.  The minimum density for a strengthless nucleus with this period and shape to be stable against centripetal loss from the equator is $\rho_n$ = 1600 kg m$^{-3}$.  Rapid rotation raises the prospect that mass loss from 331P was aided and/or triggered by a  rotational instability.

\item  The debris trail consists of centimeter sized and larger particles, having ejection speeds in the few to 10 cm s$^{-1}$ range,  dispersed by radiation pressure, and distributed according to a power-law relation with differential  index $q = 3.7\pm0.1$ to $q =  4.1\pm0.2$ (radii $\gtrsim$ 1 cm).   

\item At least 19 discrete fragments are embedded within the debris trail.  The largest, 331P-A, is a 100 m scale body having  a 1.3 to 1.5 magnitude (factor of four) lightcurve range, deep and V-shaped lightcurve minima and a period near 9 hours.  These properties are consistent with a contact binary structure.

\item Discrete fragments ejected from 331P contain  about 0.01 of the mass of the primary, and this mass is dominated by the largest fragments in the distribution.   The small fractional mass loss   indicates a weak ``mass shedding''  instability of loose surface materials, not a deep structural instability in which the primary is disrupted.

\end{enumerate}

\acknowledgments
We thank Pedro Lacerda, Jane Luu and the anonymous referee for comments on the manuscript.  Based on observations made with the NASA/ESA Hubble Space Telescope, obtained from the data archive at the Space Telescope Science Institute. STScI is operated by the Association of Universities for Research in Astronomy, Inc. under NASA contract NAS 5-26555.  Support for this work was provided by NASA through grant number AR-16618 from the Space Telescope Science Institute, which is operated by AURA, Inc., under NASA contract NAS 5-26555.


%% in parentheses, after the keyword, but they will not be verified.

{\it Facilities:}  \facility{HST}.

\clearpage


\clearpage

\begin{deluxetable}{lccrrrrrrr}
\tabletypesize{\scriptsize}
%\rotate
\tablecaption{Observing Geometry 
\label{geometry}}
\tablewidth{0pt}
\tablehead{\colhead{UT Date \& Time} & \colhead{N\tablenotemark{a}}  & \colhead{Exp\tablenotemark{b}} & \colhead{$\nu$\tablenotemark{c}}  & \colhead{$r_H$\tablenotemark{d}} & \colhead{$\Delta$\tablenotemark{e}}  & \colhead{$\alpha$\tablenotemark{f}} & \colhead{$\theta_{- \odot}$\tablenotemark{g}} & \colhead{$\theta_{-V}$\tablenotemark{h}} & \colhead{$\delta_{\oplus}$\tablenotemark{i}}   }

\startdata

2015 Dec 25 09:33-14:58 	&  24 & 368 & 39.3 		& 2.906 & 2.011 & 9.7 & 64.7 & 269.0 & 3.9 \\
2015 Dec 28 01:09-03:23 	&  12 & 368 & 39.8 		& 2.907 & 2.030 & 10.5 & 66.2 & 269.0 & 4.0 \\
2016 Feb 13 03:47-09:16 &  24 & 368 & 49.3 		& 2.920 & 2.573 & 19.4 & 78.3 & 269.6 & 3.6 \\
2017 Feb 13 02:19-09:16 & 24 & 438 & 118.8		& 3.059 & 2.111 & 6.2 & 331.0 & 286.2 & 4.3  \\
2017 Mar 08 16:13-23:00 & 25 & 438 & 123.0 		& 3.068 & 2.109 & 5.8 & 70.4 & 286.7 & 3.4 \\
2018 May 17 13:59-16:10 & 10 & 438 & 197.4 		& 3.122 & 2.113 & 1.6 & 169.0 & 274.5 & -1.5 \\
2018 May 26 09:19-16:17 & 20 & 438 &  198.9 		& 3.121 & 2.126 & 4.3 & 125.4 & 274.9 & -2.2 \\
2018 Jul 03 09:33-16:31	& 20 & 453 & 205.3 		& 3.115 & 2.402 & 15.2 & 110.0 & 275.2 & -3.7 \\

\enddata

%% Text for table notes should follow after the \enddata but before
%% the \end{deluxetable}. Make sure there is at least one \tablenotemark
%% in the table for each \tablenotetext.

\tablenotetext{a}{Number of images obtained on each date}
\tablenotetext{b}{Representative integration time per image  }
\tablenotetext{c}{True anomaly, in degrees}
\tablenotetext{d}{Heliocentric distance, in AU }
\tablenotetext{e}{Geocentric distance, in AU }
\tablenotetext{f}{Phase angle, in degrees }
\tablenotetext{g}{Position angle of projected anti-solar direction, in degrees }
\tablenotetext{h}{Position angle of negative heliocentric velocity vector, in degrees}
\tablenotetext{i}{Angle from orbital plane, in degrees}

\end{deluxetable}

\clearpage

\begin{deluxetable}{lccccccc}
%\tabletypesize{\scriptsize}
%\rotate

\tablecaption{Fragment Photometry\tablenotemark{a} 
\label{photometry}}
\tablewidth{0pt}

\tablehead{ \colhead{Object} & $H_5$\tablenotemark{a} &  $H_{m}$\tablenotemark{b}  & $N$\tablenotemark{c} & $r_n$\tablenotemark{d} [km] }
\startdata
%2017 Jun 28 	& 0.44			& 21.59/9.45/2.5 & 20.80/8.66/5.2 & 20.04/7.90/10.4	& 19.34/7.20/19.8 & 18.83/6.69/32 & 18.63/6.49/38  \\
Primary	&	17.92	$\pm$	0.03	&	17.96	&	7	&	0.771	$\pm$	0.011	\\
A	&	22.31	$\pm$	0.04	&	22.25	&	7	&	0.110	$\pm$	0.002	\\
B	&	23.35	$\pm$	0.08	&	23.31	&	7	&	0.067	$\pm$	0.002	\\
C	&	23.13	$\pm$	0.06	&	23.13	&	7	&	0.073	$\pm$	0.002	\\
D	&	23.87	$\pm$	0.17	&	23.95	&	3	&	0.050	$\pm$	0.004	\\
E	&	23.18	$\pm$	0.12	&	23.18	&	2	&	0.071	$\pm$	0.004	\\
F	&	24.12	$\pm$	0.07	&	24.12	&	2	&	0.046	$\pm$	0.001	\\
G	&	23.75	$\pm$	0.13	&	23.75	&	2	&	0.055	$\pm$	0.003	\\
H	&	24.03	$\pm$	0.32	&	24.03	&	2	&	0.048	$\pm$	0.007	\\
I	&	23.76	$\pm$	0.09	&	23.77	&	4	&	0.054	$\pm$	0.002	\\
J	&	23.54	$\pm$	0.10	&	23.55	&	6	&	0.060	$\pm$	0.003	\\
K	&	23.77	$\pm$	0.15	&	23.77	&	2	&	0.054	$\pm$	0.004	\\
L	&	23.52	$\pm$	0.10	&	23.52	&	2	&	0.061	$\pm$	0.003	\\
M	&	23.58	$\pm$	0.10	&	23.61	&	5	&	0.059	$\pm$	0.003	\\
N	&	23.44	$\pm$	0.11	&	23.45	&	4	&	0.063	$\pm$	0.003	\\
O	&	23.85	$\pm$	0.13	&	23.76	&	5	&	0.055	$\pm$	0.003	\\
P	& 	23.66	$\pm$	0.12	&	23.66	&	2	&	0.057	$\pm$	0.003	\\
Q	&	23.67	$\pm$	0.10	&	23.62	&	3	&	0.058	$\pm$	0.003	\\
R	&	24.43	$\pm$	0.31	&	24.43	&	1 	&	0.040	$\pm$	0.006	\\
S	&	23.60	$\pm$	0.20	&	23.70	&	1 	&	0.057	$\pm$	0.006	\\

\enddata

%% Text for table notes should follow after the \enddata but before
%% the \end{deluxetable}. Make sure there is at least one \tablenotemark
%% in the table for each \tablenotetext.
\tablenotetext{a}{Mean value of the absolute magnitude within a 5 pixel (0.2\arcsec)~radius aperture, with the standard deviation on the mean}
\tablenotetext{b}{Median value of the absolute magnitude obtained in the same way}
\tablenotetext{c}{Number of images used}
\tablenotetext{d}{Effective radius, in km, computed from $H_m$, Equation (\ref{C}) and $r_n = (C/\pi)^{1/2}$}

\end{deluxetable}

\clearpage
%%%%%%%%%%%%%%%%%%%%%%%%%%%%%%%%%%%%%%%%%
%%%%%%%%%%%%%%%%%%%%%%%%%%%%%%%%%%%%%%%%%
%%%%%%%%%%%%%%%%%%%%%%%%%%%%%%%%%%%%%%%%%

\begin{deluxetable}{lccrrrrrrr}
\tabletypesize{\scriptsize}
%\rotate
\tablecaption{Breakups Compared
\label{breakups}}
\tablewidth{0pt}
\tablehead{\colhead{Object} & \colhead{$a$\tablenotemark{a}}  & \colhead{$e$\tablenotemark{b}} & \colhead{$i$\tablenotemark{c}}  & \colhead{$r_n$\tablenotemark{d}} &  \colhead{$P$\tablenotemark{e}}  & \colhead{$\Delta M/M$\tablenotemark{f}}  & \colhead{$N$\tablenotemark{g}} & \colhead{Reference\tablenotemark{h}}    }

\startdata
P/2013 R3  & 3.033 & 0.273 & 0.9 & 0.4? & ? & $\sim$1 & 12 & J14, J17 \\
331P  & 3.005 & 0.042 & 9.7 & 0.8$\pm$0.1 & 3.26$\pm$0.01 & 3$\times10^{-3}$ & 19 &  This Work\\
311P & 2.189 & 0.116 & 5.0 & 0.19$\pm$0.03 & $\ge$5.4 &  $10^{-4}$ to $10^{-3}$ & 0 & J13, H14, J18 \\
6478 Gault   & 2.306 & 0.193 & 22.8 & 2.0$\pm$0.4 & 2.55$\pm$0.10 & $\gtrsim3\times10^{-6}$ & 0 & J19, S19, L21 \\

\enddata

%% Text for table notes should follow after the \enddata but before
%% the \end{deluxetable}. Make sure there is at least one \tablenotemark
%% in the table for each \tablenotetext.

\tablenotetext{a}{Semimajor axis}
\tablenotetext{b}{Orbital eccentricity}
\tablenotetext{c}{Orbital inclination, degree}
\tablenotetext{d}{Effective nucleus radius, km}
\tablenotetext{e}{Nucleus rotation period, hour}
\tablenotetext{f}{Fractional mass ejected}
\tablenotetext{g}{Number of fragments detected}
\tablenotetext{h}{J13 = Jewitt et al.~2013, J14 = Jewitt et al.~2014b, J17 = Jewitt et al.~2017, Jewitt et al.~2018, J19 = Jewitt et al.~2019, H14 = Hainaut et al.~2014, L21 = Luu et al.~2021, S19 = Sanchez et al.~2019}

\end{deluxetable}

%%%%%%%%%%%%%%%%%%%%%%%%%%%%%%%%%%%%%%%%%
%%%%%%%%%%%%%%%%%%%%%%%%%%%%%%%%%%%%%%%%%
%%%%%%%%%%%%%%%%%%%%%%%%%%%%%%%%%%%%%%%%%

%

\clearpage
%%%%%%%%%%%%%%%%%%%%%%%%%%%%%%%%%%%%%%%%%
%%%%%%%%%%%%%%%%%%%%%%%%%%%%%%%%%%%%%%%%%
%%%%%%%%%%%%%%%%%%%%%%%%%%%%%%%%%%%%%%%%%

\clearpage
%%%%%%%%%%%%%%%%%%%%%%%%%%%%%%%%%%%%%%%%%
%%%%%%%%%%%%%%%%%%%%%%%%%%%%%%%%%%%%%%%%%
%%%%%%%%%%%%%%%%%%%%%%%%%%%%%%%%%%%%%%%%%

\clearpage
%%%%%%%%%%%%%%%%%%%%%%%%%%%%%%%%%%%%%%%%%
%%%%%%%%%%%%%%%%%%%%%%%%%%%%%%%%%%%%%%%%%
%%%%%%%%%%%%%%%%%%%%%%%%%%%%%%%%%%%%%%%%%

\clearpage
%%%%%%%%%%%%%%%%%%%%%%%%%%%%%%%%%%%%%%%%%
%%%%%%%%%%%%%%%%%%%%%%%%%%%%%%%%%%%%%%%%%
%%%%%%%%%%%%%%%%%%%%%%%%%%%%%%%%%%%%%%%%%

\begin{figure}
\epsscale{0.95}
%\plotone{SB.pdf}
%\plotone{images.pdf}
%\plotone{primary_phased.pdf}
\plotone{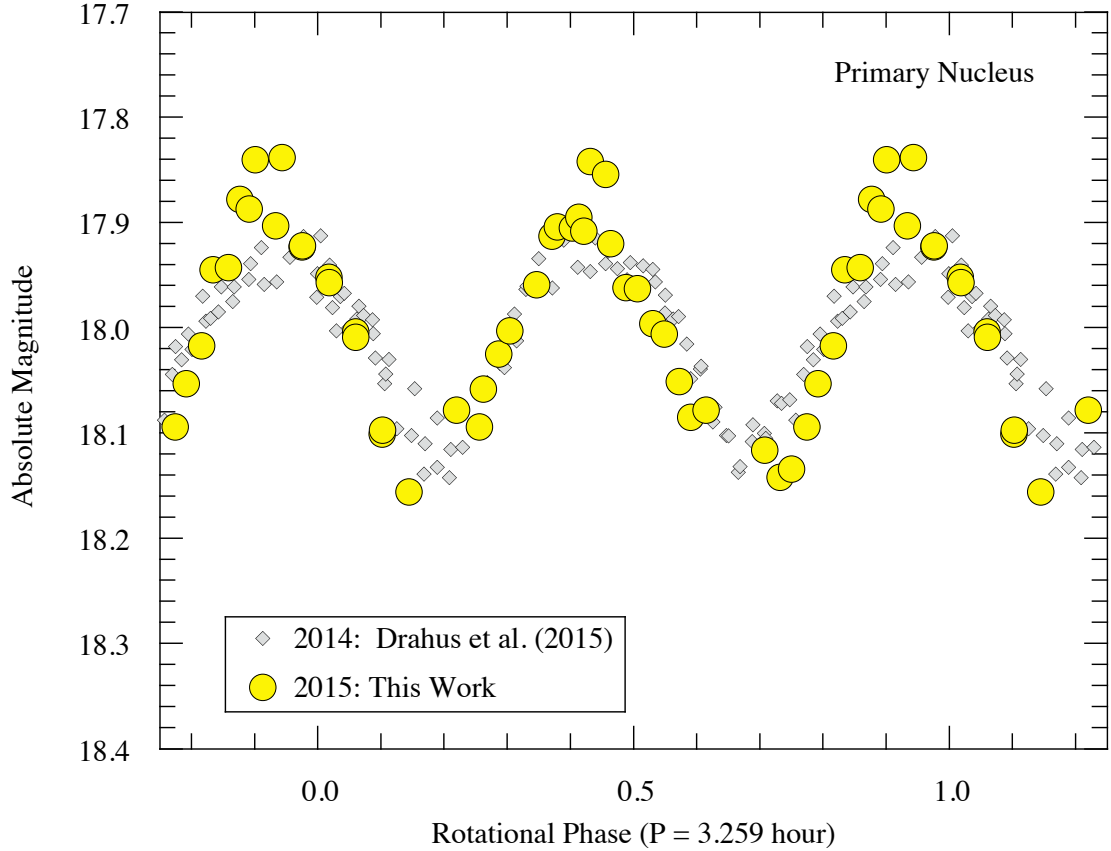}

\caption{Primary lightcurve from HST data taken UT 2015 December 25 and 28 (yellow circles) compared with data from Drahus et al.~(2015) taken in 2014.  The Drahus et al.~data have been shifted in phase and absolute magnitude to obtain the best match.  The slight differences between the lightcurves  reflect a change in the observing geometry between 2014 and 2015 as well as greater dilution of the lightcurve by near-nucleus dust in the 2014 data (see Figure \ref{absmags}).  \label{primary}}
\end{figure}

\clearpage

\begin{figure}
\epsscale{0.90}
%\plotone{SB.pdf}
%\plotone{images.pdf}
%\plotone{tail_pa.pdf}
\plotone{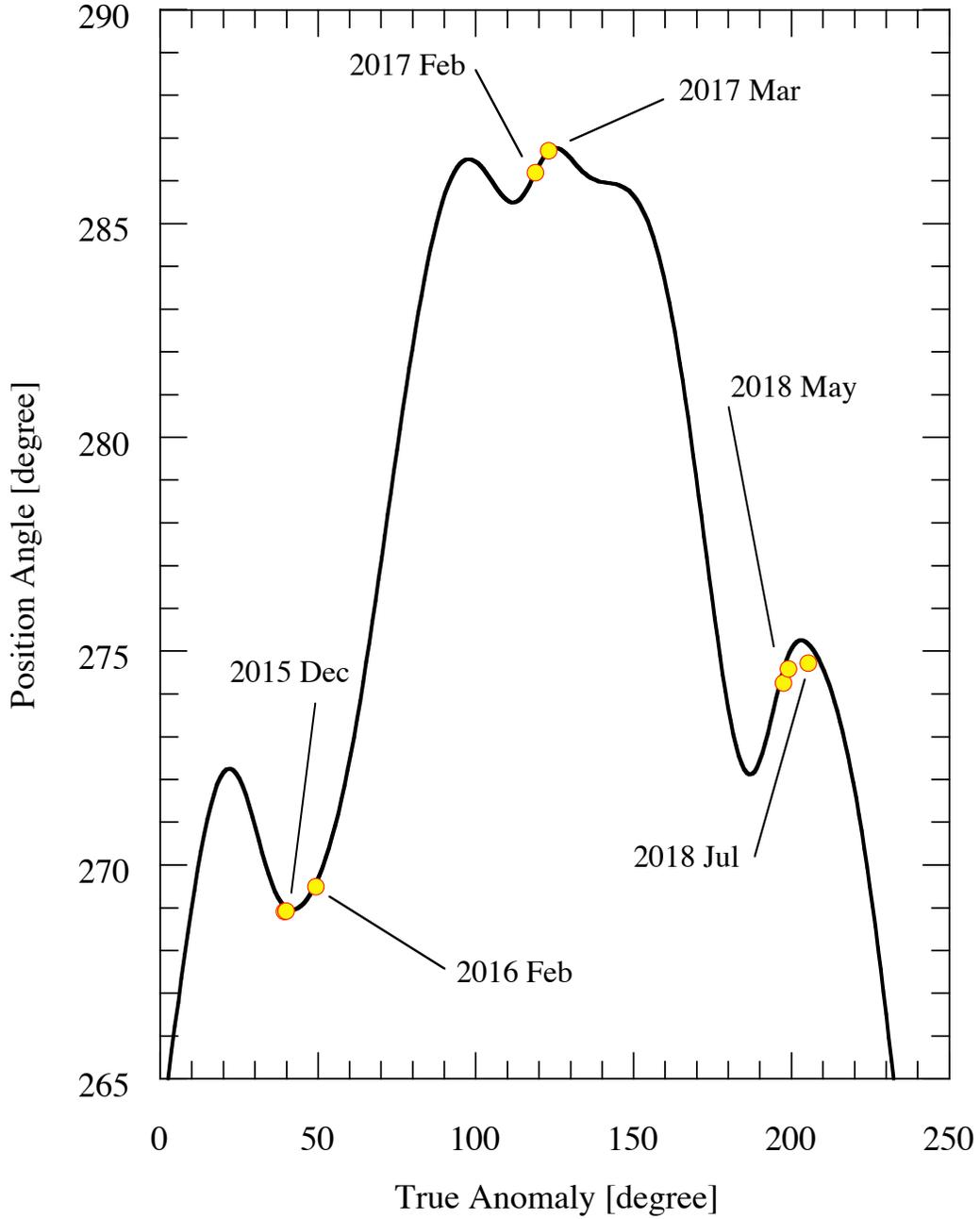}

\caption{Position angle of the trail (yellow circles) relative to the nucleus as a function of the true anomaly.  Statistical error bars on the measurements are smaller than the plot symbols.  The solid black line shows the position angle of the projected negative velocity vector. \label{PA}}
\end{figure}

\clearpage

\begin{figure}
\epsscale{0.90}
%\plotone{sb_perp.pdf}
\plotone{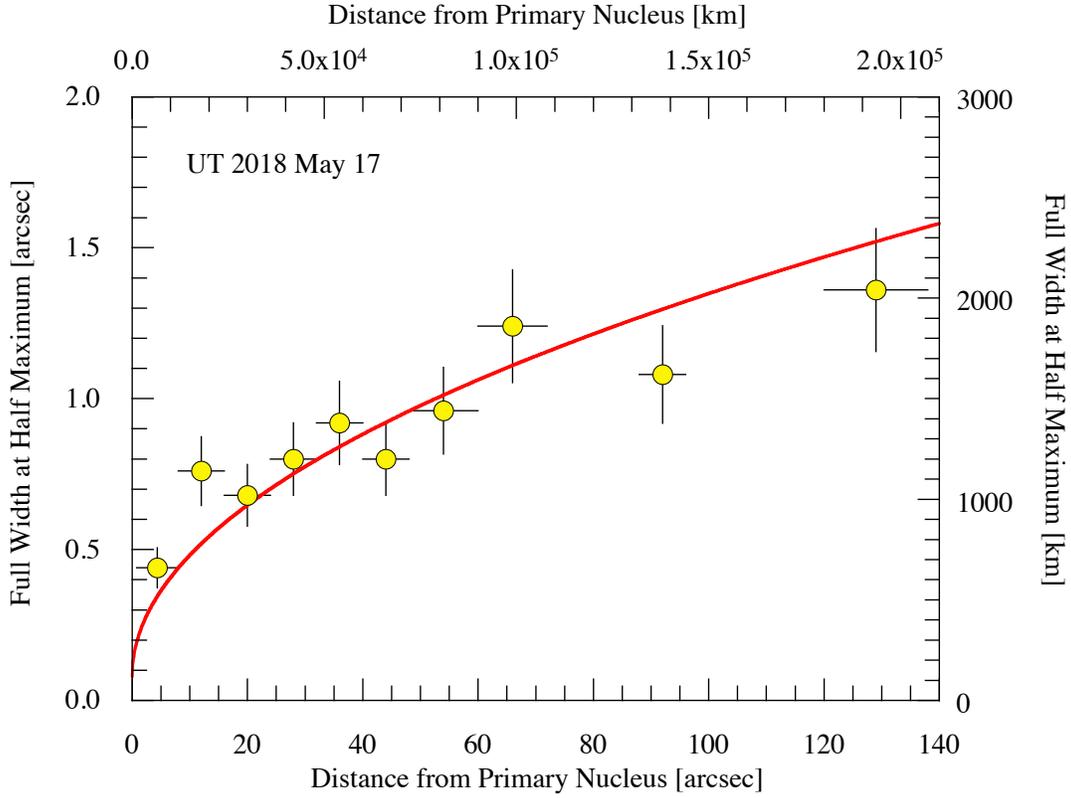}
\caption{Full width at half maximum, $W_{\perp}$, measured perpendicular to the trail axis as a function of distance from the primary, $\ell_T$.   Representative error bars equal to 15\% of the FWHM are shown. A weighted least-squares fit to a $W_{\perp} \propto \ell_T^{1/2}$ power law is shown as a red line.  \label{sb_perp}}
\end{figure}

\clearpage

\begin{figure}
\epsscale{0.80}
%\plotone{SB.pdf}
%\plotone{images.pdf}
%\plotone{sb2.pdf}
\plotone{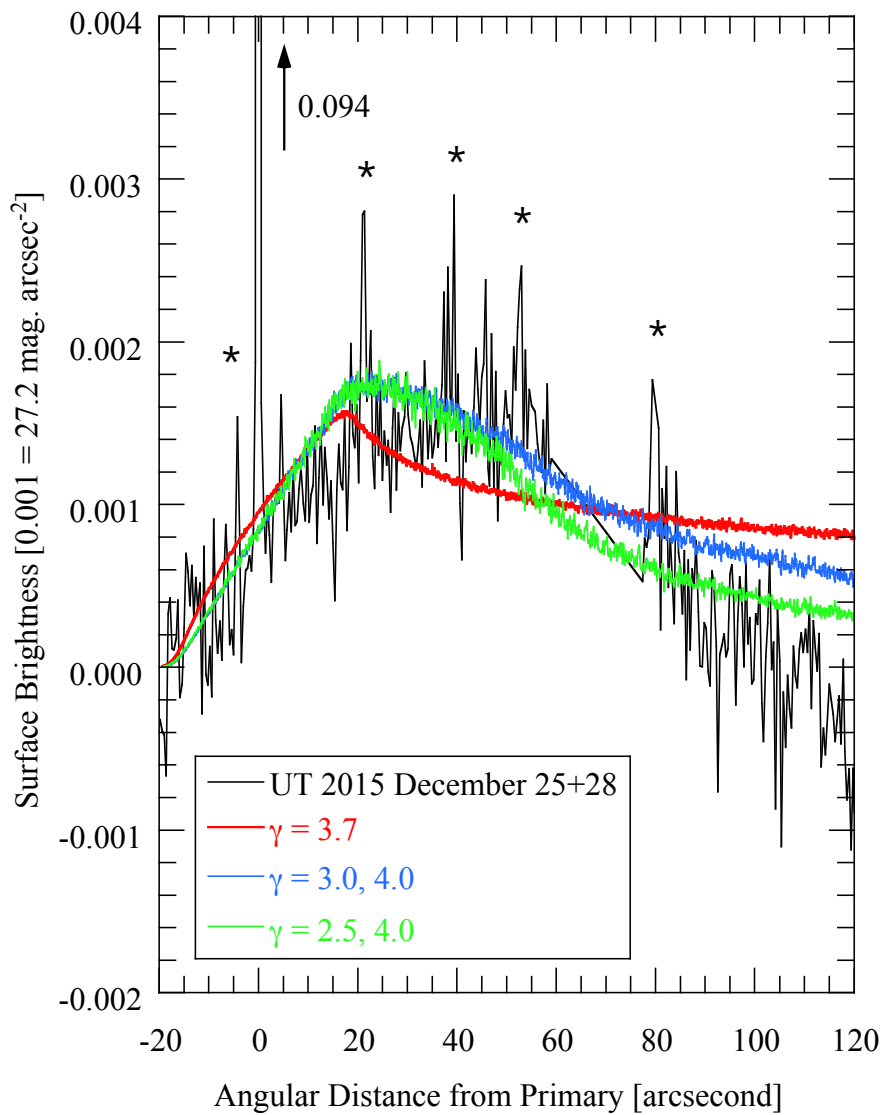}

\caption{Measured V-band surface brightness  along the trail axis (black line). The primary nucleus is off-scale, with a core surface brightness 0.094 units.   Lines show models for the brightness produced by radiation pressure sweeping of (red line) a $q$ = 3.7 power law, (blue line) a broken power law with $q$ = 3.0 and 4.0 and a break at $\beta_{crit} = 10^{-5}$ and (green line) a broken power law with $q$ = 2.5 and 4.0 and the same break $\beta_{crit}$.   Interfering stars and galaxies are marked ``$*$''.  \label{sb}}
\end{figure}

\clearpage

\begin{figure}
\epsscale{0.90}
%\plotone{SB.pdf}
%\plotone{images.pdf}
%\plotone{4objects.pdf}
\plotone{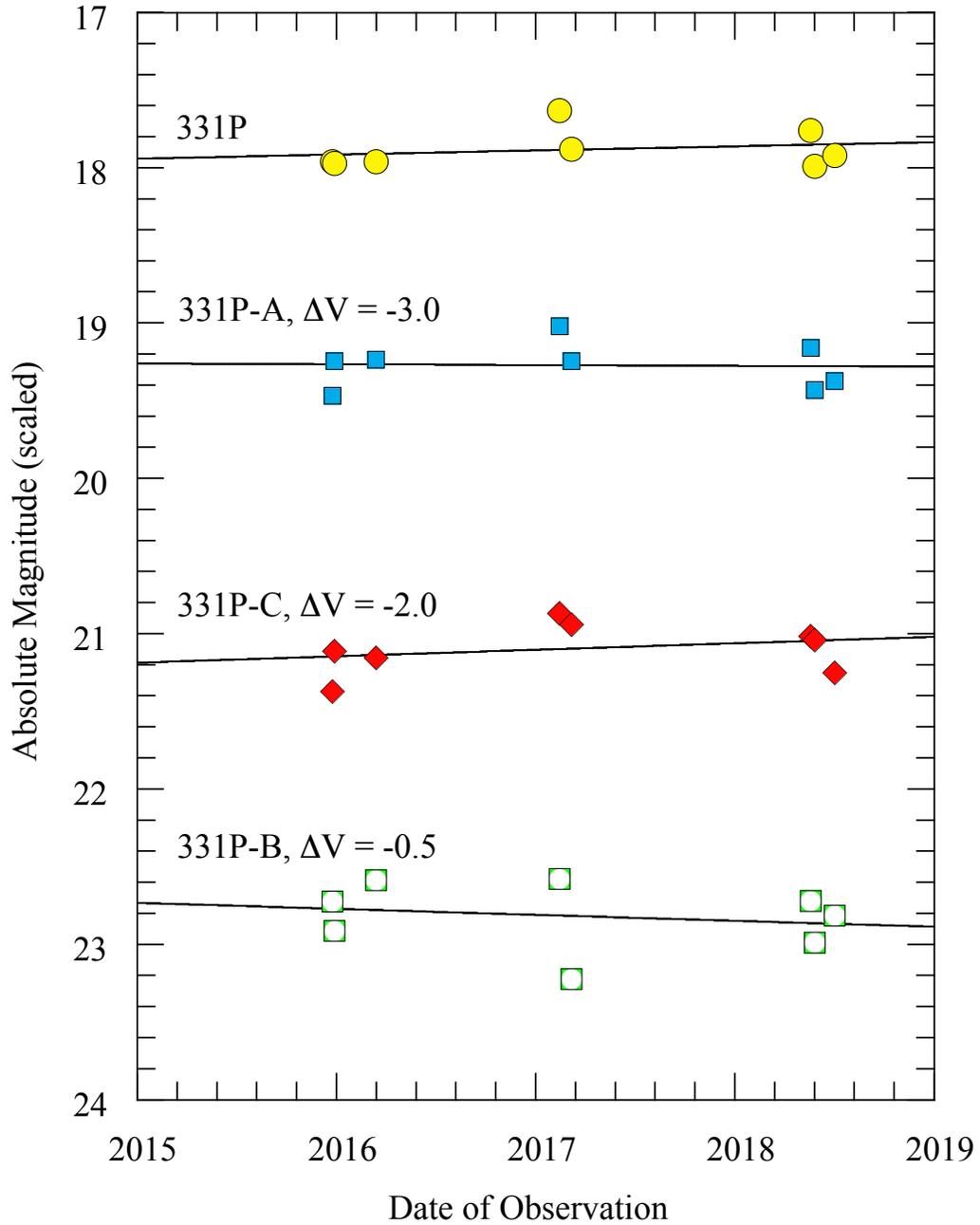}

\caption{HST photometry of 331P and fragments 331P-A, -B and -C as a function of the date of observation.  The fragment magnitudes have been vertically offset by $\Delta V$ for clarity, as indicated.  There is no evidence for a secular decline in the brightness of any object.  \label{4objects}}
\end{figure}

\clearpage

\begin{figure}
\epsscale{0.90}
%\plotone{SB.pdf}
%\plotone{images.pdf}
%\plotone{secondary_light.pdf}
\plotone{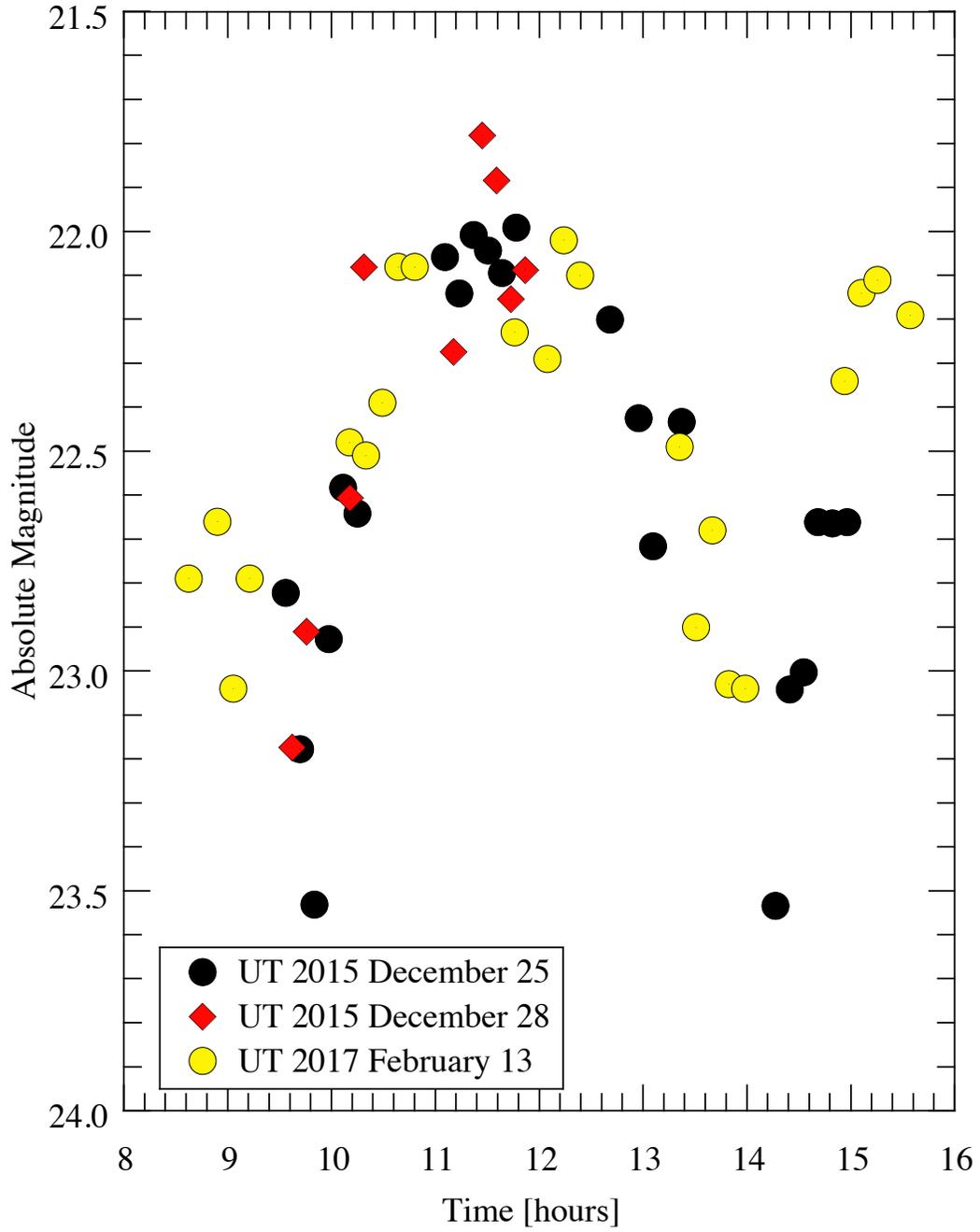}

\caption{Lightcurve of 331P-A in data from UT 2015 December 25 and 28 and UT 2017 February 13.  The times have been shifted arbitrarily to obtain the best fit.  \label{secondary}}
\end{figure}

\clearpage

\begin{figure}
\epsscale{0.90}
%\plotone{SB.pdf}
%\plotone{images.pdf}
%\plotone{size_plot.pdf}
\plotone{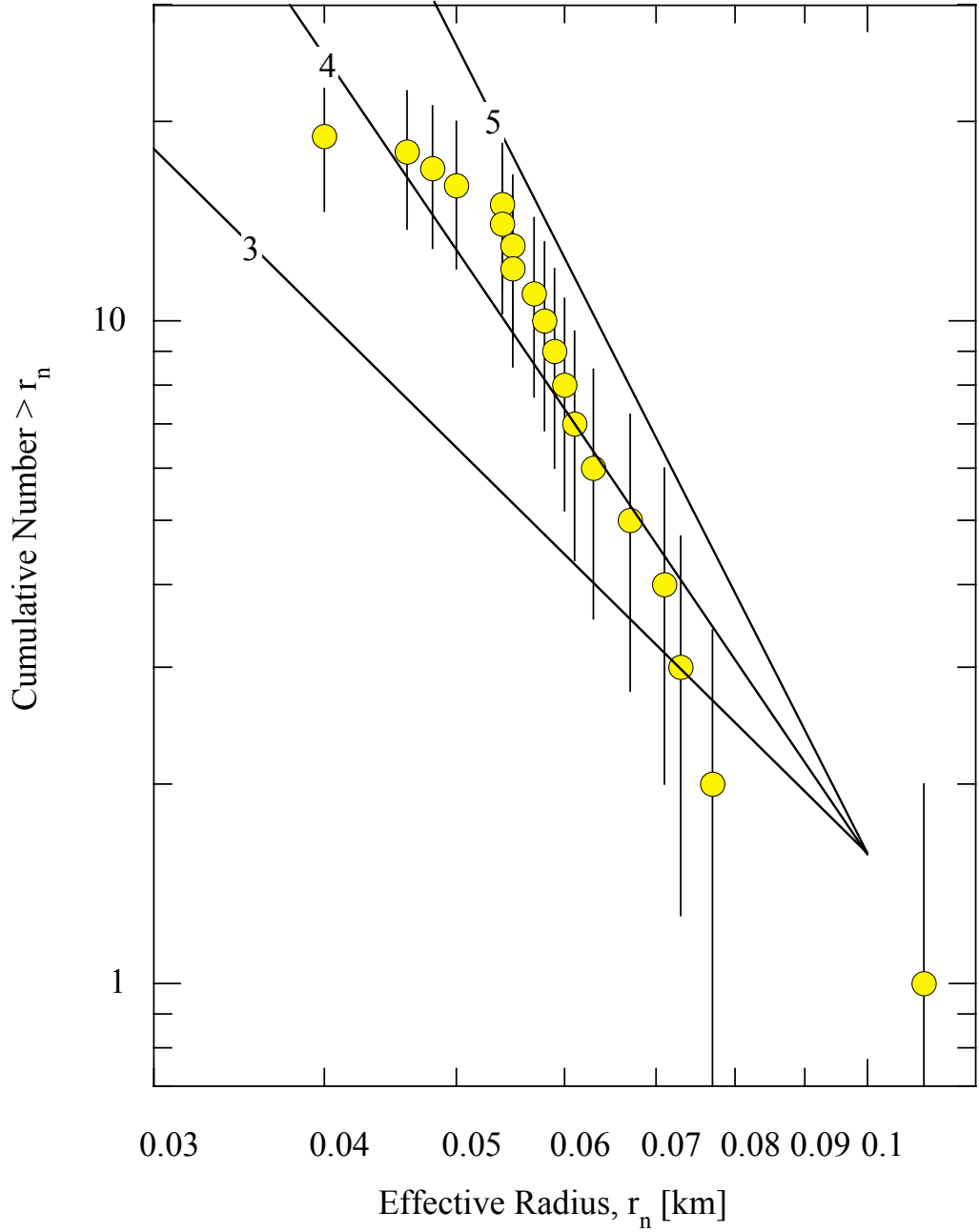}

\caption{Cumulative distribution of fragment radii.  Lines indicate differential power-law indices 3, 4 and 5, as labeled.  \label{size_plot}}
\end{figure}

\clearpage

\begin{figure}
\epsscale{0.95}
%\plotone{SB.pdf}
%\plotone{images.pdf}
%\plotone{schematic}
\plotone{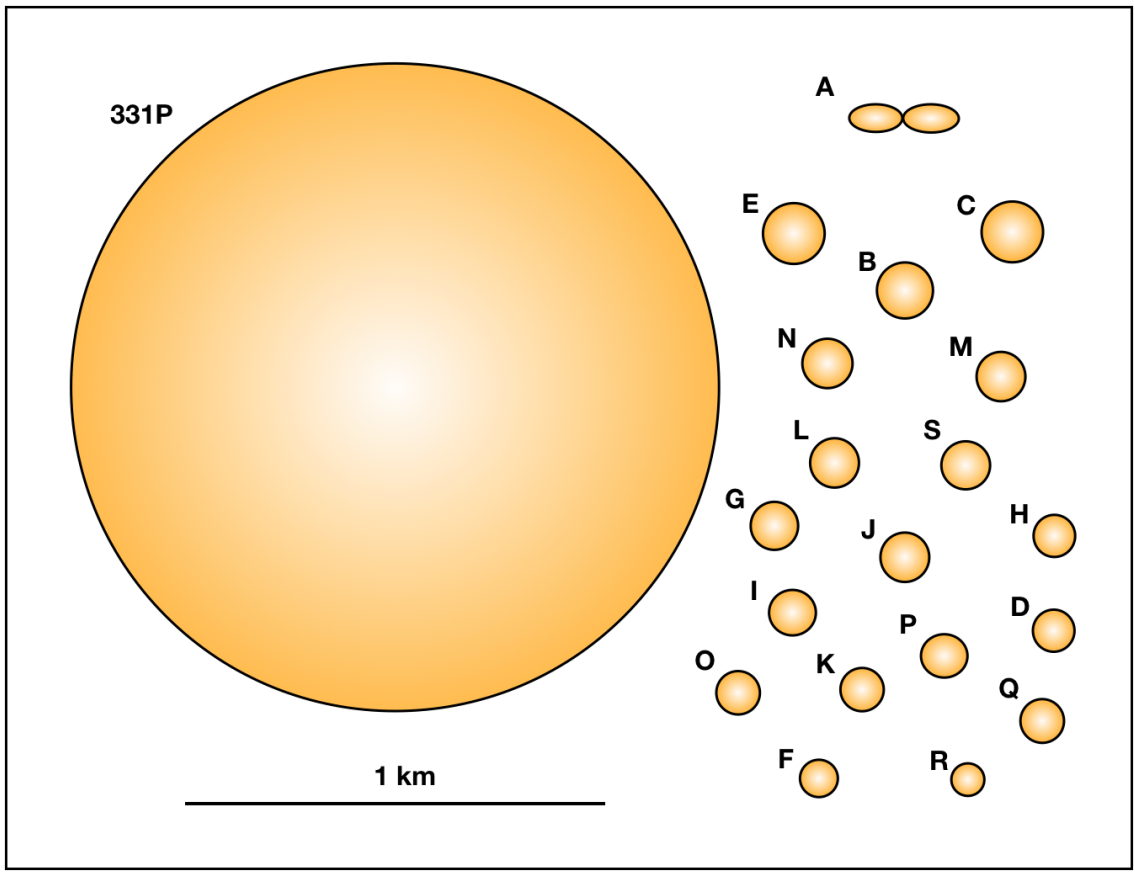}

\caption{Schematic diagram to show the relative sizes of 331P and its fragments.  \label{schematic}}
\end{figure}

\clearpage

\begin{figure}
\epsscale{0.65}
%\plotone{SB.pdf}
%\plotone{images.pdf}
%\plotone{compared.pdf}
\plotone{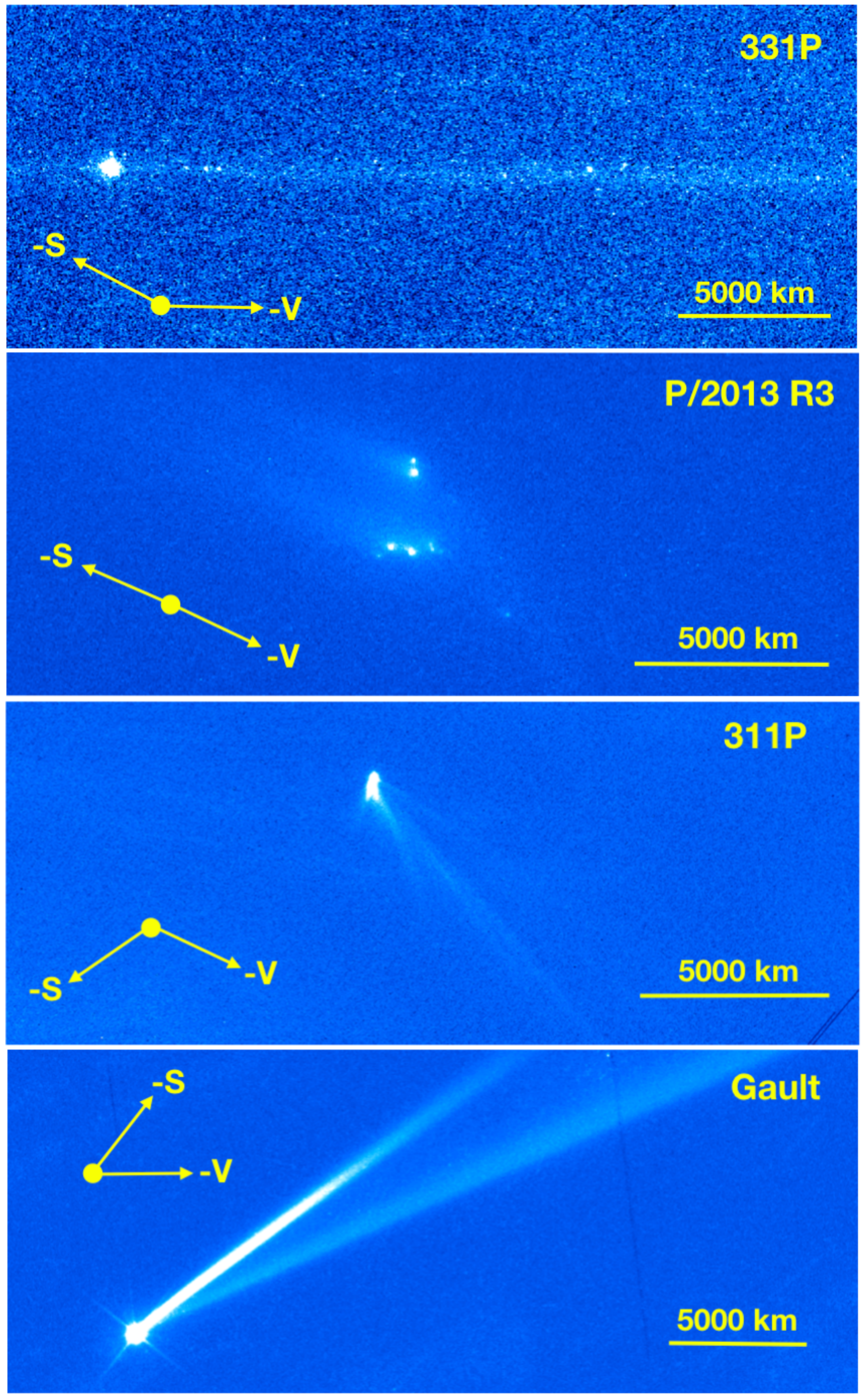}

\caption{Four asteroids inferred to have been activated by rotation instabilities, as observed by HST.  331P is from this work, P/2013 R3 from Jewitt et al.~(2014b), 311P from Jewitt et al.~(2013) and Gault from Kleyna et al.~(2019).  Each panel has North to the top, East to the left and is shown with a 5000 km scale bar and arrows indicating the projected anti-solar vector ($-S$) and the negative heliocentric velocity vector ($-V$).  \label{compared}}
\end{figure}

%\clearpage
%
%\begin{figure}
%\epsscale{0.8}
%%\plotone{SB.pdf}
%\plotone{radial_profiles.pdf}
%\caption{Dust cross-section as a function of aperture size and observation date. Points show the measured values listed in Table~\ref{photometry}. Lines represent linear fits to the two innermost measurements, assuming a steady state coma.
%\label{radial_profiles}}
%\end{figure}

\end{document}